\documentclass[12pt,preprint]{aastex}
\usepackage{natbib}
\usepackage{threeparttable}
\usepackage{graphicx}
\usepackage{epstopdf}
\usepackage{rotating}
\usepackage{lscape}
\shortauthors{Kryukova et al.}

\bibpunct{(}{)}{,}{a}{}{;}
\bibpunct{(}{)}{,}{a}{}{;}

\begin{document}

\title{The Dependence of Protostellar Luminosity on Environment in the Cygnus-X Star-Forming Complex}

\author{E. Kryukova\altaffilmark{1}, S.~T. Megeath\altaffilmark{1}, J. L. Hora\altaffilmark{2}, R.~A. Gutermuth\altaffilmark{3}, S. Bontemps\altaffilmark{4,5}, K. Kraemer\altaffilmark{6}, M. Hennemann\altaffilmark{7}, N. Schneider\altaffilmark{4,5}, Howard~A. Smith\altaffilmark{2}, F. Motte\altaffilmark{7}}

\altaffiltext{1}{Ritter Astrophysical Observatory, Department of Physics and Astronomy, University of Toledo, Toledo, OH; megeath@physics.utoledo.edu}
\altaffiltext{2}{Harvard-Smithsonian Center for Astrophysics, Cambridge, MA}
\altaffiltext{3}{Department of Astronomy, University of Massachusetts, Amherst, MA}
\altaffiltext{4}{Univ. Bordeaux, LAB, UMR 5804, F-33270, Floirac, France.}
\altaffiltext{5}{CNRS, LAB, UMR 5804, F-33270, Floirac, France}
\altaffiltext{6}{Institute for Scientific Research, Boston College, 140 Commonwealth Avenue, Chestnut Hill, MA 02467, USA}
\altaffiltext{7}{Laboratoire AIM, CEA/IRFU - CNRS/INSU -  Universit\'e Paris Diderot, Service d'Astrophysique, B\^at. 709, CEA-Saclay, 91191 Gif-sur-Yvette Cedex, France}

\begin{abstract}
The Cygnus-X star-forming complex is one of the most active regions of low and high mass star formation within 2 kpc of the Sun.  Using mid-infrared photometry from the IRAC and MIPS {\it Spitzer} Cygnus-X Legacy Survey, we have identified over 1800 protostar candidates.  We compare the protostellar luminosity functions of two regions within Cygnus-X: CygX-South and CygX-North. These two clouds show distinctly different morphologies suggestive of dissimilar star-forming environments.  We find the luminosity functions of these two regions are statistically different.  Furthermore, we compare the luminosity functions of protostars found in regions of high and low stellar density within Cygnus-X and find that the luminosity function in regions of high stellar density is biased to higher luminosities. In total, these observations provide further evidence that the luminosities of protostars depend on their natal environment.  We discuss the implications this dependence has for the star formation process.  
\end{abstract}

\keywords{infrared: stars, stars: protostars, stars: formation, stars: luminosity function}

\section{Introduction}
\label{sec:intro}

Cygnus-X is a highly obscured area of the sky which encompasses some of the most active regions of star formation in our galaxy known to date.  It contains multiple OB associations, including the massive Cygnus OB2 association which hosts 65 known  O stars, hundreds of B stars, and tens of thousands of low mass stars \citep[e.g.][]{2012A&A...543A.101C,2010ApJ...713..871W,2000A&A...360..539K,1991AJ....101.1408M}, as well as numerous well-studied sites of ongoing massive star-formation including S106, W75, and DR 21 and at least 40 massive protostars \citep{2007A&A...474..873S,2007A&A...476.1243M}.  Cygnus-X  is found near the galactic tangent at $l=80^{o}$, and hence kinematic velocities cannot be used to estimate distances of individual regions.  Due to the large number of HII regions and star-forming regions within Cygnus-X, it has been long debated whether Cygnus-X is a superposition of multiple clouds along the line of sight or a single star formation complex \citep[][ and references therein]{2008hsf1.book...36R}.  A recent study of the velocity structure in the CO gas by \cite{2006A&A...458..855S} supported the view that Cygnus-X is mostly a single complex surrounding the Cygnus OB2 association at a distance of 1.7 kpc. Further evidence for the ``single complex" view has been found in a maser parallax studies of five regions in Cygnus-X by \cite{2012A&A...539A..79R}; they found that 4 out of 5 regions are at a common distance of 1.4 kpc.  (We adopt the 1.4 kpc distance from the maser parallax study, which is considered more reliable than the 1.7 kpc distance derived from main sequence fitting). 

Due to its proximity and concentration of massive star-forming regions, Cygnus-X has emerged as a key target in studies of high mass star formation, the interaction between molecular clouds and OB associations, and studies of star and planet formation in the OB association environment \citep[see for example][]{2012ApJ...744...86Z,2007MNRAS.374...29D,2012ApJ...746L..21W,2012arXiv1208.0211W}.  With a mass of $7.3 \times 10^5$~$M_{\odot}$ \citep{2006A&A...458..855S}, about seven times that of Orion molecular clouds \citep{2005A&A...430..523W}, Cygnus-X provides us with a nearby region that is comparable in mass, spatial scale and luminosity to the  giant star-forming regions identified in external galaxies.

Large scale surveys of the Cygnus-X region at IR and millimeter wavelengths show extensive populations of young stars and protostars embedded within the molecular gas.  \cite{2007A&A...476.1243M} surveyed the two primary molecular gas clouds in Cygnus-X: the CygX-North cloud, which contains DR21 and W75 among other regions, and the CygX-South cloud, which shows multi-parsec long pillars interacting with the Cyg OB2 association (Schneider et al 2012). They discovered 129 massive cores with masses ranging from 4-950~$M_{\odot}$, 42 of which contained massive protostars. The {\em Spitzer} Cygnus-X Program (P.I. Hora) mapped a 24 $deg^{2}$ field toward the complex with the IRAC and MIPS instruments onboard the {\it Spitzer Space Telescope}.  \cite{2010ApJ...720..679B} studied the CygX-North cloud using the {\it Spitzer} survey data and found 670 protostars and 7,249 pre-main sequence stars with disks in this region alone.  The {\it Spitzer} survey has been supplemented with near-IR photometry from the 2MASS all sky survey and the UKIDSS galactic plane survey resulting in eight band photometry from 1.2 to 24~$\mu$m for many sources \citep{2008MNRAS.391..136L}. Most recently, Herschel mapped Cygnus-X in the far-IR with the PACS and SPIRE instruments, revealing extended filamentary structures \citep{2012A&A...543L...3H}.

In this paper, we make the first systematic survey of {\it Spitzer}-identified protostars in the Cygnus-X region.  We employ the methodology of  \cite{2012AJ....144...31K} (hereafter, KMG12), who studied a sample of {\it Spitzer}-identified protostars in nine nearby ($<$ 1 kpc) star-forming regions. KMG12 used IRAC and {\it Spitzer} 24~$\mu$m photometry to identify protostars and to estimate their bolometric luminosity. After constructing the protostellar luminosity functions for each of the nine clouds, they found significant differences between the protostellar luminosity functions of clouds which form high mass stars and clouds which do not.  In addition, by comparing the luminosity functions of protostars found in regions of high surface densities of young stellar objects (YSOs) to those in regions of low YSO surface densities, KMG12 found that the luminosity function in the denser regions  extended to higher luminosities than those for lower density regions.  This final result suggested that regions with higher YSO densities preferentially form more luminous protostars, and potentially more massive stars.

Although more distant than the regions studied in KMG12, the Cygnus-X cloud complex provides  a much richer sample of protostars in a more extreme range of environments.  In this paper, we employ a slightly modified version of the analysis developed by KMG12 to identify protostars, determine their mid-infrared (mid-IR) and bolometric luminosities, construct protostellar luminosity functions and correct the luminosity functions for contamination. We also perform an analysis of the spatially varying incompleteness of the protostars.  Taking into account this incompleteness, we compare different regions within Cygnus-X to search for variations in the protostellar luminosity function between clouds within the complex.  Furthermore, we compare  regions of high and low YSO density to test whether the protostellar luminosity function extends to higher luminosities in dense regions, as found for Orion by KMG12.  With these results, we argue that the protostellar luminosity function in Cygnus-X depends on the environment in which the protostars form.

\section{Protostar Identification}
\subsection{Near- \& Mid-Infrared Photometry}
\label{sec:photometry}
The Cygnus-X complex was the target of the Cygnus-X {\it Spitzer} Legacy Survey\footnote{From {\em http://irsa.ipac.caltech.edu/data/SPITZER/Cygnus-X/} }, which mapped 24 deg$^{2}$ of the complex with the Infrared Array Camera \citep[IRAC,][]{2004ApJS..154...10F} and Multiband Imaging Photometer for {\it Spitzer} \citep[MIPS,][]{2004ApJS..154...25R} instruments on board the {\it Spitzer Space Telescope}\footnote{Data delivery document can be found at http://irsa.ipac.caltech.edu/data/SPITZER/Cygnus-X/docs/CygnusDataDelivery1.pdf}.  We used IRAC aperture photometry and MIPS 24~$\mu$m point-spread function (PSF) photometry from Data Release 1 Point Source Catalog with no alteration.  The PSF photometry extraction at the MIPS 24~$\mu$m wavelength is used to minimize source confusion due to nebulosity.  However, the brightest sources are saturated in the 24 $\mu$m image and the photometry is not available for these sources.  Thus, our study excludes the sources with the brightest $m_{24}$ magnitudes in Cygnus-X.

The photometry in the five {\it Spitzer} bands is augmented by photometry at near-IR J, H, and K-bands.  This photometry is preferentially from the UKIRT Infrared Deep Sky Survey (UKIDSS) \citep{2008MNRAS.391..136L}, but for sources for which UKIDSS data do not exist, we use available photometry from the Two Micron All-Sky Survey \citep[2MASS,][]{2003tmc..book.....C,2006AJ....131.1163S}.

\subsection{Protostar Candidate Selection}
\label{sec:criteria}
Mid-infrared colors can be used to identify infrared young stellar objects (YSOs) by the excess infrared emission radiated by their dusty disks and envelopes  \citep[e.g.,][]{1994ApJ...434..614G,2001A&A...372..173B,2004ApJS..154..363A,2004ApJS..154..367M,2004ApJS..154..379M,2004ApJ...617.1177W,2006ApJS..167..256R,2007ApJ...663.1149H, 2007ApJ...669..493W,2009ApJS..184...18G}.  We use the criteria of KMG12 to identify protostars in the Cygnus-X point source catalog.  These criteria are based on identification of protostars with flat and rising spectral energy distributions (SEDs) between 4.5 and 24 $\mu$m.  We use the spectral index $\alpha$ = $\delta \log(\lambda F_{\lambda})/\delta (\log(\lambda)$) calculated between 3.6 and 24 $\mu$m, to determine this slope.  Following the approach of KMG12, we search for protostars with in-falling protostellar envelopes by identifying sources with both flat SEDs (-0.3 $<$ $\alpha$$<$ 0.3), which correspond to flat spectrum protostars, and rising SEDs (0.3 $>$ $\alpha$), which correspond to Class 0/I protostars \citep{1984ApJ...287..610L,1994ApJ...434..330C,2007ApJ...669..493W,2009ApJ...692..973E}.  KMG12 implemented this approach by converting the spectral indicies into a series of color criteria involving the four IRAC bands and the MIPS 24~$\mu$m band.  They require that all protostars show a detection in the 24~$\mu$m band.

While most of our protostar selection criteria is the same as that of KMG12, we adjust one set of criteria.  Normally protostar candidates are selected in part by their red [3.6] - [4.5] color; however, due to the large amount of scattered light in the 3.6 and 4.5 $\mu$m bands, some protostars have a very red [4.5] - [24] color but a [3.6] - [4.5] color between 0.7 and 0 $mag$.  KMG12 selected these sources by the criteria in their Equation 3. However, due to the distance and bright nebulosity of Cygnus-X, nebulous knots may be mis-identified as protostars by the same criteria.  These knots are distinguished by their bright PAH emission at 3.6 $\mu$m and weak emission at 4.5 $\mu$m, which results in a [3.6] - [4.5] color less than 0 $mag$.  We subject the sources which fit Equation 3 of KMG12 to the following additional condition:

\begin{eqnarray}
& [3.6] - [4.5] \ge 0
\label{eqn:add}
\end{eqnarray}

\noindent which eliminates 15 likely nebulous knots with very blue [3.6] - [4.5] colors due to their bright 3.6 $\mu$m PAH emission.  In comparison, only one of the protostar candidates found in the sample of nearby clouds by KMG12 satisfies the criteria shown in their Equation 3, but fails the criterion in Equation \ref{eqn:add} above.  

\begin{figure}
\plotone{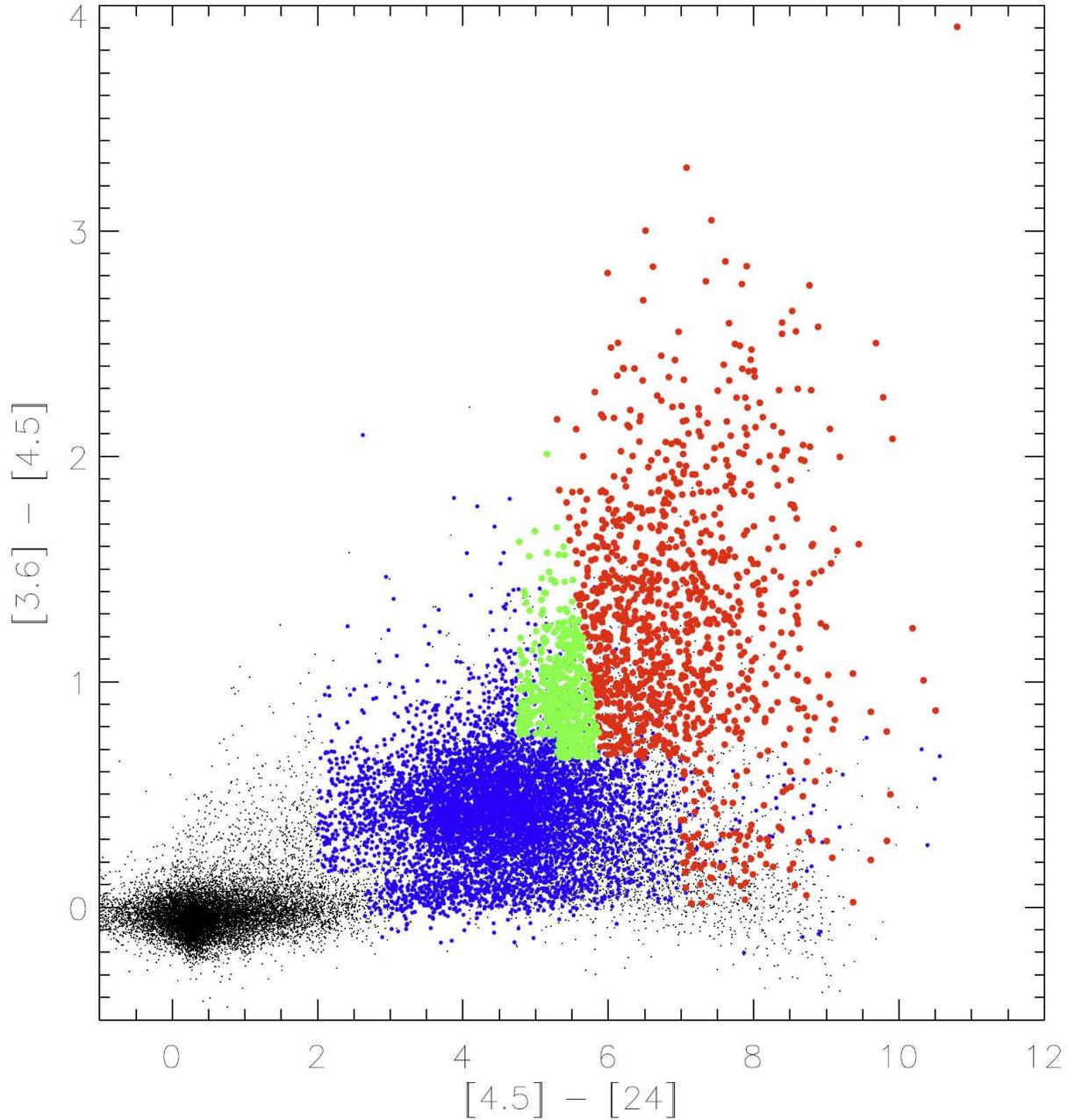}
\caption{Color-color diagram for {\it Spitzer}-identified protostars.  The rising spectrum protostars are shown in red and flat spectrum protostars are in green.  The sources we identify as Class II sources are shown in blue.  The remaining sources, shown as black dots, are stars without disks, AGN, and star-forming galaxies. }
\label{fig:c_c}
\end{figure}

\begin{figure}
\plotone{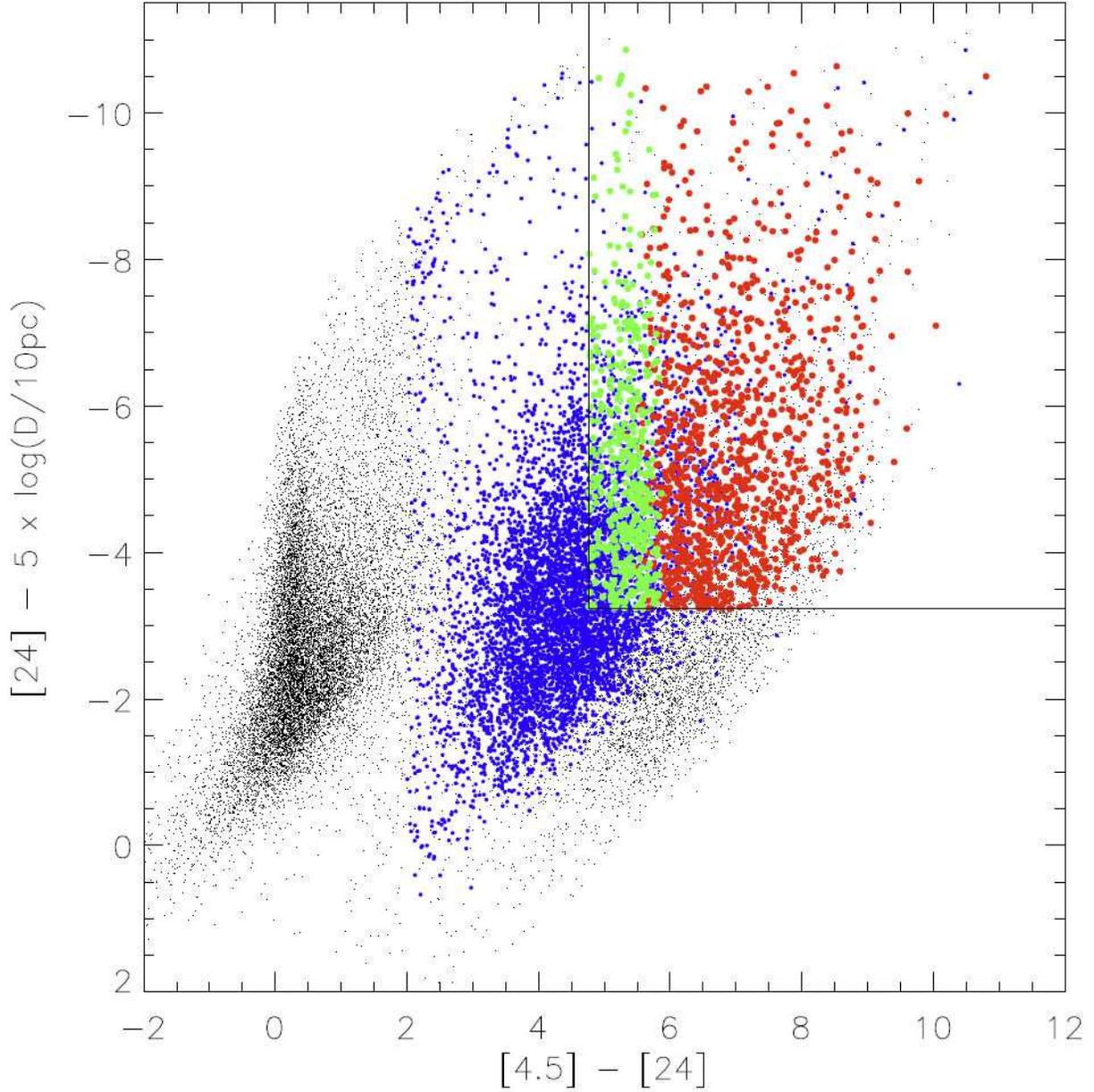}
\caption{Color-magnitude diagrams for {\it Spitzer} identified protostars.  The colored symbols are the same as in Figure \ref{fig:c_c}.  The solid lines show the $m_{24}$ cutoff and the [4.5] - [24] color cutoff.  The value of $m_{24}$ is corrected for distance but not reddening.  Galaxies comprise a distinct clump of fainter (at $m_{24}$) and highly red sources, located near [4.5] - [24] = 6.}
\label{fig:c_cm}
\end{figure}

The color-color and color-magnitude diagrams we use to identify protostar candidates are shown in Figure \ref{fig:c_c} and Figure \ref{fig:c_cm}.  Sources that satisfy any of the criteria in Table \ref{table:proto_crit} are considered protostar candidates.  
We give the number of rising and flat SED protostars, distinguished by $\alpha$ $>$ 0.3 as rising SED protostars and 0.3 $>$ $\alpha$ $>$ -0.3 as flat SED protostars, in Table \ref{table:properties}.

The spatial distribution of these sources is overlaid on the extinction map of the region shown in Figure \ref{fig:av_map}.   Note that most of the protostar candidates are concentrated in high extinction regions.  To further minimize contamination, we require that protostar candidates must lie in locations where the  visual extinction measured toward background stars is $>$ 3, as determined from the extinction map shown in Figure \ref{fig:av_map}.  This extinction map, covering the {\it Spitzer}-mapped region, is derived using {\em AvMAP} from Sylvain Bontemps \cite[private communication; see][]{2011A&A...529A...1S}.  The resulting coverage area with $A_{V} > 3$ is 21.75 $\deg^{2}$.  For comparison, Figure  \ref{fig:CO_map} shows the $^{13}$CO J = 1$\rightarrow$0 velocity integrated intensity map also from \cite{2011A&A...529A...1S}.  By requiring a minimum $A_{V}$, we limit the area in which we search for protostars to regions which are known to contain  molecular gas. Since protostars will be concentrated in regions  with high column densities of gas while contaminating objects will be more evenly spread throughout the sky, this has the advantage of reducing the amount of contamination. 

\begin{threeparttable}
\caption{Number of protostar candidates identified by selection criteria}
\begin{tabular}{c l c c}
\hline\hline
Equation & Selection Criteria Colors & Number & Number \\
& & & $A_{V} >  3$ \\
\hline
(1)\tnote{a} , (2)\tnote{a} & [4.5]-[24], [3.6]-[4.5] & 1781 & 1611 \\
(3)\tnote{a} , (1)\tnote{b} & [4.5]-[24], [5.8]-[8.0], [3.6]-[4.5] & 127 & 111 \\
(4)\tnote{a,c}  &[5.8]-[24], [3.6]-[5.8] & 179 & 143 \\
(5)\tnote{a,d}  & [8.0]-[24], [3.6]-[8.0] & 170 & 142 \\
\hline
\end{tabular}
\begin{tablenotes}
\item[a]{Criteria given from KMG12.}
\item[b]{Criteria from KMG12 that were modified in this work}
\item[c]{Only for sources without 4.5~$\mu$m detections.}
\item[d]{Only for sources without 4.5~and 5.8~$\mu$m detections.}

\end{tablenotes}
\label{table:proto_crit}
\end{threeparttable}

\vskip 0.3 in
All protostar candidates selected using these criteria must also satisfy a 24~$\mu$m magnitude cutoff.  This cutoff is imposed to eliminate galaxies with colors similar to YSOs, but which have fainter $m_{24}$ mag \citep{2009AJ....137.4072M}.  We determine the $m_{24}$ magnitude cutoff using the method of KMG12. In this method, we first determine the spatial density of galaxies with colors similar to protostars using the point source catalog from the {\it Spitzer} Wide-Area Infrared Extragalactic Survey Legacy Program \citep[SWIRE,][]{2003PASP..115..897L}\footnote{We used the 4.2~$deg^2$ Elais-N2 field from {\em http://swire.ipac.caltech.edu/swire/swire.html}}.  We compare the distribution of $m_{24}$ for the Cygnus-X protostar candidates in the region where $A_{V} >  3$  and for the SWIRE sample which satisfies the protostar candidate selection criteria.  The SWIRE histogram is scaled up to account for the larger angular extent of the $A_{V} > 3$ region.  We do not apply reddening to the SWIRE sample making our cutoff more conservative than necessary.  However, for an $A_{V} > 3$ the extinction at 24~$\mu$m is only $A_{24}$ $>$ 0.3 \citep{2009ApJ...699.1866C}, hence extinction may not have a large effect on the cutoff.  The $m_{24}$ cutoff of $m_{24}$ = 7.5 $mag$ is shown in Figures \ref{fig:c_cm} and \ref{fig:24_cutoff}.  Below this $m_{24}$ cutoff we find 2007 protostar candidates projected on regions with $A_{V} >  3$. In comparison, there are 77.7 SWIRE galaxies that fit our protostar selection criteria colors after accounting for the larger angular extent of the $A_{V} >3$ area, resulting in a contamination rate of 4.2\%. Another potential source of contamination are HII regions and reflection nebulosity in the molecular cloud complex.  Because we require that the sources in our catalog be point sources in the IRAC and MIPS bands, these would have to be very ultracompact HII regions or very small reflection nebulae with sizes $\le 7000$~A.U., the  spatial resolution at the distance of Cygnus-X.  Future, higher resolution near-IR observations may provide an effective means of further eliminating such compact HII regions and nebulae from our sample of protostars.   The number of protostar candidates within and throughout the $A_{V} > 3$ regions are given in Table \ref{table:properties}. Note that since we do not consider the effect of extinction on the magnitudes of the SWIRE galaxy, we are probably overestimating the number of contaminating galaxies.  The near and mid-infrared photometry of the 2007 protostar candidates projected on $A_{V} > 3$ are listed in Table~4.

\begin{threeparttable}
\caption{Cygnus-X YSOs}
\begin{tabular}{c c c c c c}
\hline\hline
&  Area & Class II\tnote{a} &  Protostars & & Contamination  \\
& [$deg^{2}$]& & Flat SED\tnote{b} & Rising SED\tnote{b} & {red/edge-on/galaxy\tnote{c,d}}\\
\hline
Cygnus-X & 21.75 & 4946 &  512/486/408.4 & 1495/1352/1169.0 & 60.8/121.7/78.0 \\
CygX-North & 3.72 & 2521 & 150/138/124.9 & 360/295/256.6 & 12.3/26.0/13.0 \\
CygX-South & 4.87 & 1829 & 120/115/101.1 & 341/320/283.1 & 8.0/25.7/17.0\\
\hline
\end{tabular}
\begin{tablenotes}
\item[a]{Class II sources in regions with $A_{V} > 3$ and de-reddened $m_{24}$ $<$ cutoff and RMEDSQ $<$ 10.41.}
\item[b]{All protostar candidates projected on $A_{V} > 3$/$A_{V} > 3$ nebulosity filtered protostars (see Section \ref{sec:RMEDSQ})/$A_{V} > 3$ nebulosity filtered protostars after contamination removal. Nebulosity filtered protostars are defined in Sec.~\ref{sec:RMEDSQ}.}
\item[c]{{Displayed are the mean numbers for the three forms of contaminations shown in the following order: reddened disk sources, edge-on disks sources, and extra-galactic sources. }}
\item[d]{{Uncertainties in these numbers are equal to the square root of the number of sources.}}
\end{tablenotes}
\label{table:properties}
\end{threeparttable}

\vskip 0.3 in

In addition, we look for more evolved pre-main sequence stars with disks, or Class II objects, to estimate the contamination in our protostar luminosity functions (Section \ref{sec:PLF}) and to measure the density of all detectable, dusty young stellar objects (Section \ref{sec:density}).  To identify such sources, we use the criteria of \cite{2009ApJS..184...18G}.  We find 254 candidate Class II sources with [4.5] - [24] $<$ 2; these sources show an excess at IRAC bands but do not have $m_{24}$ excesses found toward Class II sources.  Since we expect that any source exhibiting an IR-excess in the IRAC bands will also exhibit an excess in the MIPS 24~$\mu$m band \citep[e.g.][]{2012AJ....144..192M}, the sources without 24~$\mu$m excesses are likely to be pure photospheric sources, such as background stars, contaminated by nebulosity.  To eliminate these sources, we require that sources satisfy an additional criterion for Class II candidate selection:

\begin{eqnarray}
& [4.5] - [24] \ge 2
\end{eqnarray}

\noindent which all Class II sources with detections at 4.5 and 24 $\mu$m must satisfy.  We eliminate these potential contaminants from our Class II sample.  The Class II sources are also shown in Figures \ref{fig:c_c} and \ref{fig:c_cm}.  Listed in Table \ref{table:properties} are the number of Class II sources we find in regions of $A_{V} > 3$ as well as the number of protostar candidates in these regions.


\subsection{Building the Protostellar Luminosity Function}
\label{sec:PLF}
To determine the protostellar luminosity function in Cygnus-X, we must first estimate bolometric luminosities for each of our protostar candidates.  Since the majority of the flux from a protostar is emitted at wavelengths longward of the photometry bands available in this study, we use the relationship between mid-IR luminosity (calculated from 1-24 $\mu$m), the SED slope, and the bolometric luminosity developed by KMG12.  This relationship was determined using a sample of low- to intermediate-luminosity protostars (0.03 $L_{\odot}$ to 19.6 $L_{\odot}$).  We calculate the mid-IR luminosity of the protostar candidates using Equation 6 from KMG12.  Once determined, we use the mid-IR luminosity and SED slope with Equations 7 and 8 from KMG12 to find the bolometric luminosity. The median slope of Cygnus-X protostar candidates is 0.75, and 90\% of Cygnus-X protostar candidates have slopes less than 1.57. 

The range of luminosity we find for the protostar candidates in Cygnus-X extends from 0.1 $L_{\odot}$ to 3370 $L_{\odot}$, with 89\% of the protostar candidates having luminosities below 19.6 $L_{\odot}$.  Objects with luminosities $> 3000$~L$_{\odot}$ are missing due to saturation in the 24~$\mu$m band.  Since the luminosity depends both on the flux in all the 3.6 through 24~$\mu$m bands and the slope of the SED, a definite saturation luminosity cannot be established.  {Future work analyzing Herschel and WISE data may establish the number of luminous sources missed in the Spitzer survey.  \cite{2007A&A...476.1243M} find the presence of 25 massive protostars in their study of the Cygnus-X; we expect the number of missing sources to be of this order of magnitude. }

The systematic uncertainties in the luminosities determined by this method are described by KMG12; they find the log of the ratio of the well-determined bolometric luminosity to the estimated bolometric luminosity has a mean value of 0.06 with a standard deviation of 0.35.  We expect that neither the systematic uncertainties nor the uncertainties from the fluxes used in determining $L_{MIR}$ should affect the gross properties of the displayed luminosity functions. Furthermore, even if systematic uncertainties affect the conversion of the measured $L_{MIR}$ into bolometric luminosities, such uncertainties would presumably affect all sources in the same way, and thereby leave the comparisons unaffected.  The bolometric luminosities of the protostar candidates are listed in Table~4.

\subsubsection{Completeness at $m_{24}$}
\label{sec:RMEDSQ}
The derived protostellar luminosity function is affected by spatially varying completeness (KMG12, Megeath et al. 2012).  Since we require of our protostars a detection in $m_{24}$, we are primarily concerned with the completeness in this band, and in comparison to the 3.6 and 4.5 $\mu$m bands the nebulosity will have the most effect at 24 $\mu$m. We follow the approach of Megeath et al. (2012) to determine the completeness as a function of the fluctuations due to nebulosity and confusion from other point sources.

The completeness in the {\it Spitzer} images is spatially varying primarily due to the bright, highly structured nebulosity found in {\it Spitzer} images of star-forming regions.  Instead of measuring the completeness independently at each position, we measure the completeness as a function of source magnitude and the RMEDSQ, a measure of the fluctuations in the signal due to nebulosity and neighboring point sources in an annulus immediately surrounding a point source.  Megeath et al. (2012) define $RMEDSQ = \sqrt{median[(S_{ij}-median(S_{ij}))^2]}$, where $S_{ij}$ is the signal in DN~s$^{-1}$~pix$^{-1}$ at a given pixel position {\it i,j}.  The median values are determined for an annulus centered on each source; this annulus typically extends from 6 to 11 pixels but is increased for more luminous sources \citep{2012AJ....144..192M}.  Thus, we first assign to each of our sources a RMEDSQ.

To measure the completeness to point sources as a function of the RMEDSQ, we create an overlay of PSFs arranged in a 3 $\times$ 3 grid and add it to the {\it Spitzer} 24 $\mu$m mosaic.  The grid is created by using the MIPS 24 $\mu$m PSF scaled to varying magnitudes, including $m_{24}$ = 3.125, 3.625, 4.125, 4.625, 5.125, 5.625, 6.125, 6.625, and 7.125, spaced 25 pixels from one another.  This grid is then repeatedly added to the Cygnus-X 24 $\mu$m image at intervals of 151 pixels across the mosaic so that the entire image is covered by the repeating grid.  The mosaic has a pixel size of 1."2.  We use {\em PhotVis} \citep{2008ApJ...674..336G} to find all the 24 $\mu$m point sources in the mosaic and perform aperture photometry on the sources.  We consider an input PSF recovered if a point source is found  within 2."5 of the input PSF's position with a magnitude within 0.25 {\em mag} of the input PSF's magnitude.  The RMEDSQ is determined around each of the input PSFs.  For each input magnitude, these sources are then binned by RMEDSQ.  Figure \ref{fig:MAD_mag} shows the fraction of recovered sources as a function of the $log(RMEDSQ)$ for each of the input magnitudes.

To mitigate the incompleteness in extended regions of bright nebulosity and extreme source crowding, we first measure the incompleteness in the region surrounding each protostar.  To do so, we use the average RMEDSQ value of YSOs (protostar candidates projected on  $A_{V} >  3$ regions and Class II sources) within a critical distance, $D_{c}$ = 0.62 pc (see Section \ref{sec:density}), of each protostar, which we denote $<$RMEDSQ$>_{D_{c}}$ value for that protostar.  Using the mean RMEDSQ value for nearby YSOs provides a more representative measurement of the typical fluctuations in the regions surrounding each protostar than using  only the RMEDSQ value for the protostar.  The $<$RMEDSQ$>_{D_{c}}$ value is listed for each protostar candidate in Table~4.

In Figure \ref{fig:rmedsq}, we show the $<$RMEDSQ$>_{D_{c}}$ value vs. bolometric luminosity for each protostar candidate in Cygnus-X, CygX-North, and CygX-South.  This figure shows that incompleteness can result in a clear bias in our determination of the luminosity functions in that luminous protostars can be found in regions with high RMEDSQ where lower luminosity protostars cannot be detected.  To eliminate this bias, we filter out sources with either high values of RMEDSQ or low values of luminosity.  We first determine a cutoff RMEDSQ value such that 90\% of the YSOs, including other protostar candidates and Class II sources within $D_{c}$ = 0.62 pc of a protostar candidate, have $<$RMEDSQ$>_{D_{c}}$ values below this cutoff; the value of the cutoff is 10.41. Using the curves from Figure \ref{fig:MAD_mag}, we interpolate between curves to find a limiting $m_{24}$ that corresponds to a 90\% completeness at the cutoff RMEDSQ value.  The resulting limiting $m_{24}$ is 5.07 $mag$. The corresponding limiting $m_{24}$ is converted to a luminosity using Equation 6 of KMG12 assuming the entire mid-IR luminosity is given by the 24 $\mu$m flux and assuming an SED slope larger than that of 90\% of the protostar candidates in Cygnus-X. The corresponding cutoff luminosity is 2.16 $L_{\odot}$.  Thus, we find that we are 90\% complete at a luminosity cutoff of $L_{cut}$ = 2.16 $L_{\odot}$ and an $<$RMEDSQ$>_{D_{c}}$ = 10.41.  Most of the protostars will be in regions that have smaller $<$RMEDSQ$>_{D_{c}}$ values and hence higher levels of completeness at 2.16 $L_{\odot}$.

The RMEDSQ cutoff and $L_{cut}$ are also shown in Figure \ref{fig:rmedsq}. There are 1838 protostar candidates with $<$RMEDSQ$>_{D_{c}}$ values less than 10.41; we refer to these protostars as the ``nebulosity filtered'' sample in the remaining text.  We only use sources below this cutoff and luminosity greater than $L_{cut}$ = 2.16 $L_{\odot}$ when comparing the luminosity functions. The regions around the sample nebulosity filtered protostars are 99.7\% complete at $L_{cut}$. 

\subsubsection{Protostellar Luminosity Functions}
In Figure \ref{fig:blum_cont}, we show the luminosity functions of the protostar candidates.  We display the luminosity functions for the entire region, the CygX-North region, and the CygX-South region, as defined by \cite{2006A&A...458..855S} (see Section 3.0).  To probe the differences between the various star-forming environments, we compare luminosity functions of protostars in different regions.  However, there are yet multiple sources of contamination which are not eradicated by the protostar selection criteria of Section \ref{sec:criteria}, which include edge-on disks \citep{2008A&A...486..245C}, highly reddened Class II sources where the extinction results in a rising SED \citep{2009ApJS..181..321E,2010ApJS..188...75M}, and galaxies with $m_{24}$ brighter than our $m_{24}$ cutoff.  

For each type of  contamination, we generate multiple realizations of their luminosity functions using the Monte Carlo method described in Section 4 of KMG12. For sources with edge-on disks, the ratio of edge-on disks to all Class II objects and the luminosities of the edge-on disks sources are determined from a sample of likely edge-on disks in the Cep OB3b cluster as described by KMG12.  To estimate the number of edge-on disks, we multiple this ratio by the number of Class II sources with $<$RMEDSQ$>_{D_{c}}$ $<$ 10.41. The highly reddened Class II sources are identified using the same technique as in KMG12, but utilizing a fiducial sample of low-AV Class II sources with $<$RMEDSQ$>_{D_{c}}$ $<$ 10.41 from Cygnus-X. These are reddened using the extinctions toward the sample of protostars and YSOs with $A_{V} > 3$ and $<$RMEDSQ$>_{D_{c}}$ $<$ 10.4 (see KMG12 for the details of this process). The method of estimating galaxy contamination is unchanged from KMG12 and utilizes the same SWIRE sample as a galaxy sample.  The methods for determining the luminosity functions of the edge-on sources and highly reddened Class II sources require us to use the number of Class II sources with $<$RMEDSQ$>_{D_{c}}$ below 10.41. 

For each Monte Carlo realization, we determine a sample of contaminants as described by KMG12.  We calculate the luminosity of each contaminant using the mid-IR luminosity and SED slope; this may not be the actual luminosity of the contaminating source, but instead the luminosity we would assign to the source given its mid-IR photometry. We then remove a source from the protostar  luminosity function which has the closest luminosity to the contaminant, requiring that the difference is less than 0.2 dex.  If no protostar candidate has such a luminosity, then no source is removed.  We repeat this process for each of the 1000 realizations.  We stress that we are not identifying protostar candidates as contamination and removing them, instead we  predict the luminosities for ensembles of contaminating sources as generated by the Monte Carlo code and remove those luminosities from the protostellar luminosity function.   The resulting mean luminosity distributions of edge-on disks, galaxies, and reddened Class II sources found over the 1000 trials of the Monte Carlo code are shown in Figure \ref{fig:blum_cont}, along with the luminosity function of the protostar candidates.   The protostellar luminosity functions with the contamination removed are shown in Figure \ref{fig:blum_sub_hists}; these have been averaged over the 1000 trials.  The luminosity functions for the total region, the CygX-North region, and the CygX-South region are shown; the above procedure was run for each of these regions independently.  The $m_{24}$ cutoff, along with the number of contaminants, protostar candidates, Class II sources with $A_{V} > 3$, and flat and rising SED protostars are listed in Table \ref{table:properties}. 

We find that the contaminants account for only 14.2\% of our total 1838 protostar candidates.  The most common type of contamination is due to edge-on disk sources.  However, because Cygnus-X is both in the Galactic plane and a more distant cloud than those studied in KMG12, there may be additional sources of Galactic contamination such as more distant star-forming regions or more evolved stars.  We have attempted to minimize that contamination by only including protostar candidates projected on regions of $A_{V} > 3$.

\section{Protostellar Environment and Luminosity}

The distribution of protostars and their relationship to the distribution of molecular gas, as delineated by the $A_{V}$ map or the $^{13}$CO velocity integrated intensity map, is shown in Figures  \ref{fig:av_map} and  \ref{fig:CO_map}.  The locations of the most luminous protostar candidates are shown  circled in red. These maps show that the Cygnus-X complex is dominated by two cloud complexes, the CygX-North and CygX-South complex.  \cite{2006A&A...458..855S} argued that CygX-North and CygX-South are at a similar distance as the Cyg OB2 association and that Cygnus-X is mostly a coherent complex rather than a superposition of clouds at varying distances. More recently, \cite{2012A&A...539A..79R} found that DR 21, DR 20, W75N, and IRAS 20290+4052, all within the northern cloud, are consistent with a distance of 1.4 kpc, but  AFGL 2591, which is located within our boundary for CygX-South, is at a distance of 3.33$\pm$0.11 kpc. However, given the evidence that CygX-South is interacting with the Cyg OB2 association, we will assume that the CygX-North and CygX-South clouds are at 1.4 kpc, although there is some contamination with more distant regions \citep{2012A&A...542L..18S}.   

The CygX-North and CygX-South molecular clouds exhibit  different morphologies in their mid-IR emission and in the distribution of protostars relative to the molecular gas.  Maps of the mid-IR emission with the position of the protostars overlaid are shown in  Figures~\ref{fig:north_map} and \ref{fig:south_map}.  We see in Figure \ref{fig:av_map} and Figure \ref{fig:CO_map} that within our chosen boundary for CygX-North, the spatial distribution of protostars tends to follow the cloud structure apparent in both the $A_{V}$ map and $^{13}$CO map. The central cloud is filamentary in structure containing rich, multi-parsec chains of protostar candidates.  These are particularly apparent in DR 21, DR 22, DR 23 and W75; the filament containing DR21 and W75 is perhaps the most active star-forming region in the CygX-North \citep{2010ApJ...720..679B,2010A&A...520A..49S}.   Although there are UV illuminated features on the side of the cloud facing Cyg OB2 \citep{2006A&A...458..855S}, the DR21/W75 filament is not associated with extended, bright IR nebulosity. The lack of nebulosity indicates that this filament is not being directly illuminated by the Cyg OB2 association.  We also do not find an excess of luminous protostar candidates on the edge of the CygX-North cloud closest to Cyg~OB2; although, the most luminous protostar candidates are primarily found in the half of the CygX-North region closest to Cyg~OB2. This suggests that much of the star formation in the CygX-North is not being directly influenced by the UV radiation and winds from Cyg~OB2.  

In CygX-South, the protostars tend to be located toward the outer parts of the cloud on both maps, with fewer protostar candidates being found toward the center of the cloud.  Compared to CygX-North, the cloud morphology in CygX-South is less dominated by filamentary structures.  The CygX-South appears to be more strongly interacting with the Cyg OB2 association and contains prominent pillars extending towards and illuminated by Cyg OB2, which is 40 $pc$ north of the heads of the pillars \citep[Figure 11,][]{2012A&A...542L..18S}.  The part of the CygX-South region facing Cyg OB2 has an abundance of protostar candidates.  In contrast, there are relatively few protostar candidates located in some of the densest regions of the $^{13}$CO map in the central and southern parts of CygX-South.  The concentration of protostars near Cyg-OB2 suggest that a wave of star formation is being driven into CygX-South by the direct compression of the cloud surfaces by Cyg OB2.  This direct compression may be result from the photoevaporation of the cloud surfaces by the UV field from Cyg OB2 \citep[e.g.][]{1997AJ....114.1106M,2010A&A...523A...6D}.  The different spatial distributions of protostars and gas in the CygX-South and CygX-North regions  are  evidence that both radiatively triggered and a more spontaneous star formation are currently operating in Cygnus-X.

\subsection{Protostellar Luminosity Functions of CygX-North and CygX-South:  Evidence for Variations}
\label{sec:n_s}
In the previous section, we showed that the spatial distribution of protostars and the morphologies of their parental clouds appear to be different in the CygX-North and CygX-South.  Here we test whether the protostellar luminosity functions of these sub-regions are similar.   

We first comment on the shape of the Cygnus-X, CygX-North, and CygX-South luminosity functions of the nebulosity filtered protostars (Figure \ref{fig:blum_sub_hists}).  None of these luminosity functions has a  peak above $L_{cut}$ = 2.16 $L_{\odot}$, but rather in each case, the luminosity functions increase toward lower luminosity, though the Cygnus-X and CygX-North luminosity functions rise more steeply than the CygX-South luminosity function.  The CygX-North luminosity function has an excess of sources near 100 $L_{\odot}$ relative to the Cygnus-X or CygX-South luminosity functions.  Above $L_{cut}$ = 2.16, the Cygnus-X luminosity function has a maximum luminosity of 3370 $L_{\odot}$ (close to the saturation limit), a median value of $6.8^{+8.4}_{-3.8}$~$L_{\odot}$, a mean value of $30.5^{+37.8}_{-16.9}$~$L_{\odot}$ and standard deviation of $144.2^{+178.6}_{-79.8}$~$L_{\odot}$.  The uncertainties for the mean and median are due to those in the determination of $L_{bol}$, where the standard deviation of the log of the luminosity integrated over the SED of sources with far-IR photometry over that estimated from $L_{MIR}$ (i.e. $log\left[{L_{bol}(SED)}/{L_{bol}(MIR)}\right]$) is $0.35$.   The CygX-North luminosity function has a maximum luminosity of $280^{+347}_{-155}$~$L_{\odot}$, a median of $7.3^{+9.0}_{-4.0}$~$L_{\odot}$, a mean of $26.6^{+33.0}_{-14.7}$~L$_{\odot}$, and standard deviation of $46.2^{+ 57.2}_{-25.6}$~$L_{\odot}$.  In CygX-South, the maximum luminosity is $276^{+342}_{-153}$~$L_{\odot}$, with a median of $7.4^{+9.2}_{-4.1}$~$L_{\odot}$, a mean of $19.0^{+23.5}_{-10.5}$~$L_{\odot}$, and standard deviation of $34.9^{+43.2}_{-19.3}$~$L_{\odot}$.  The CygX-North and CygX-South regions have similar numbers of protostars,  with CygX-North containing 157 nebulosity filtered protostars and CygX-South containing 169 nebulosity filtered protostars with $L_{cut}$ $>$ 2.61 $L_{\odot}$.

A statistical comparison using a two sample Kolmogorov-Smirnov (K-S) test allows us to determine the probability that any two luminosity functions are from the same parent distribution.  To take into account uncertainties in the correction for contamination, we use an ensemble of 1000 distributions of each luminosity function, each of which is a realization from the Monte Carlo method used to estimate contamination. We perform the K-S test 1000 times between realization 1 from luminosity function A with realization 1 from luminosity function B, and realization 2 from luminosity function A with realization 2 from luminosity function B, and so on (see KMG12).  We then take the median probability from the 1000 realizations.  Only nebulosity filtered sources above $L_{cut}$ are used in the test.  

We first compare the luminosity function of the entire Cygnus-X sample with both the individual CygX-North and CygX-South luminosity functions.   The median probabilities from the K-S test comparing the CygX-North and CygX-South luminosity function to the entire Cygnus-X cloud luminosity function are listed in Table \ref{table:probs}.  These results indicate that the CygX-North luminosity function is not drawn from the same parent distribution as the entire Cygnus-X cloud luminosity function.  In comparison, the Cygnus-X luminosity function and CygX-South luminosity function are likely drawn from the same parent distribution. Comparing the CygX-North and CygX-South luminosity functions, we find that the protostellar luminosity function of CygX-North has a median probability $log(prob)$ = -7.47 of being from the same parent distribution as the luminosity function of the protostars in CygX-South (Table \ref{table:probs}).  If we increase $L_{cut}$ to 3~L$_{\odot}$, the probability decreases to $log(prob)$ = -9.4 that  CygX-North and CygX-South are drawn from the same distribution.  

Could differences we observe be an artifact resulting from the superposition of multiple star-forming regions at different distances?  Our determination of the protostellar luminosity adopts a single distance for each of our samples; the presence of multiple distances may result in a shift of the luminosity function or a broader spread of luminosities.  However, \cite{2006A&A...458..855S} argue on the basis of their $^{13}$CO data that the CygX-North and CygX-South are coherent cloud regions and that they are both associated with Cygnus-X.  The degree of contamination from larger distances is likely small, particularly if we limit ourselves to the CygX-North and CygX-South regions. The maser parallaxes of \cite{2012A&A...539A..79R} show only one out of five regions is a background region located at a larger distance of 3 kpc. This more distant region, AFGL 2591, is spatially enclosed in the CygX-South region, but is at a distance approximately twice that of the CygX-South region \citep{2006A&A...458..855S}.  We do not include this source in our protostar candidate list because it is saturated at $m_{24}$, and thus it does not affect the CygX-South luminosity function.  

Furthermore, a small difference in distances for CygX-North and CygX-South is unlikely to affect our comparison.  Even if CygX-North and CygX-South are on the opposite sides of the Cyg OB2 association, they would be separated by 100~pc, resulting in a small shift in $log$(L [$L_{\odot}$]) of 0.06 \citep{2006A&A...458..855S}.  Thus, because the evidence favors the CygX-North and CygX-South regions as containing coherent clouds  at nearly the same distance, we conclude that the differences in the luminosity functions between these regions are very likely to be real.   Over the entire Cygnus-X region, there may be a higher degree of contamination from protostars at different distances; this may potentially contribute to the dissimilarity between the Cygnus-X and CygX-North luminosity functions. Hence, we are not as confident of differences between the CygX-North and the Cygnus-X regions.  

\begin{threeparttable}
\caption{K-S Test probabilities}
\begin{tabular}{c c c c c c}
\hline\hline
Region &  CygX-North & CygX-South & High/Low Density & HMSFC\tnote{a} & LMSFC\tnote{b} \\
\hline
Cygnus-X & -9.93\tnote{c} & 0.9085 & -3.19\tnote{c} & 0.6343 & 0.0200   \\
CygX-North  & -  & -7.47\tnote{c} & -5.42\tnote{c} & -5.88\tnote{c} & -4.33\tnote{c}  \\
CygX-South &  -7.47\tnote{c} & -  & 0.0135 & 0.8123 & 0.0111  \\
\hline
\end{tabular}
\begin{tablenotes}
\item[a]{HMSFC are high mass star-forming clouds, see Sec.~3.3.}
\item[b]{LMSFC are low mass star-forming clouds, see Sec.~3.3.}
\item[c]{Given as $log(prob)$.}
\end{tablenotes}
\label{table:probs}
\end{threeparttable}

\subsection{Protostellar Luminosity and Stellar Density}
\label{sec:density}
KMG12 found a dependence of the protostellar luminosity function on the density of YSOs in the Orion molecular clouds, with the luminosity function being biased to higher luminosities in regions of high stellar density. We now repeat this analysis for the Cygnus-X region to see if the same dependence is apparent.  To sort protostars by the local surface density of YSOs, we use the nearest-neighbor distance.  This distance is found to the 4th nearest YSO,  which we designate ``nn4 distance'' following the approach of KMG12.  We include both the nebulosity filtered protostar candidates projected on $A_{V} > 3$ regions and Class II sources without $A_{V}$ or RMEDSQ cutoffs as YSOs. Figure~\ref{fig:nns_all} shows the distribution of bolometric luminosity vs. nn4 distance for the entire cloud sample, CygX-North sample, and CygX-South sample. Similar to what KMG12 found for nearby high mass star-forming clouds, we find that in each sample the maximum nn4 distance of the protostar candidates tends to decrease with increasing luminosity (except for the most luminous protostar candidate in the Cygnus-X sample). This trend is particularly pronounced in the CygX-North region.  At low luminosities near our completeness limit, $L_{cut}$ = 2.16 $L_{\odot}$, we find protostar candidates at both large and small nn4 distances. The spatial distribution of the protostars in these two environments are shown in Figures \ref{fig:av_map} and \ref{fig:CO_map}, with the blue and green symbols marking the positions of protostars in low  and high stellar density environments, respectively.  Note this analysis does not account for the saturated protostars missing from our sample which will be more luminous than the sample of observed protostars.

We now seek to compare luminosity functions of protostars in high and low stellar density environments.  To determine if a protostar candidate is in a high or low stellar density environment, we use the median nn4 distance, denoted $D_{c}$, for the Cygnus-X cloud nebulosity filtered sample. This provides an equal number of protostar candidates in high and low stellar density regions in Cygnus-X, which is optimal for statistical comparisons between the two samples. We find $D_{c}$ = 0.62 pc, which is equivalent to a surface density of 4.15 YSOs $pc^{-2}$. In comparison, this number is less dense than the definitions of a cluster given by \cite{2000AJ....120.3139C}, \cite{2009ApJS..184...18G}, and \cite{2010MNRAS.409L..54B}, even though \cite{2000AJ....120.3139C} and \cite{2009ApJS..184...18G} find that approximately half of YSOs are found in clusters.  The $D_{c}$ used for Cygnus-X is also larger than the $D_{c}$  used for nearby molecular clouds by KMG12 (0.08 to 0.4 pc).  However, the Cygnus-X region is more distant than the regions in the above studies and hence is more incomplete to lower mass Class II objects and protostars.  For this reason, a larger value of $D_{c}$ is appropriate for Cygnus-X.

In Figure \ref{fig:MAD_nns}, we show the $log(<$RMEDSQ$>_{D_{c}})$ deviation as a function of nn4 distance, with the sources having luminosities above 10 $L_{\odot}$ marked.  We find the most luminous protostars tend to be in high stellar density environments with high values of $<$RMEDSQ$>_{D_{c}}$.  These plots also show that the regions where ${\rm nn4} < D_{c}$ have preferentially higher $<$RMEDSQ$>_{D_{c}}$ values.  This demonstrates why we must only consider nebulosity filtered protostars even though many of the higher luminosity sources are excluded from the analysis.   By ignoring regions where the $<$RMEDSQ$>_{D_{c}}$ is above our adopted threshold, our sample of protostars in the remaining regions is 90\% complete at our luminosity cutoff, $L_{cut}$ = 2.61 $L_{\odot}$.

We note that although there are equal number of protostars in high and low stellar density regions for the entire sample, that is not true for the individual clouds.  CygX-North has a notably larger percentage of nebulosity filtered protostar candidates with luminosities above $L_{cut}$  in high stellar density environments (67.3\%) as does the CygX-South region (60.4\%).  Furthermore, the percentage of sources in high stellar density regions increase for more luminous sources. In CygX-North, 77.2\% of the nebulosity filtered protostar candidates with luminosity greater than 10 $L_{\odot}$ are in high stellar density environments,  in CygX-South this number is 69.6\%, and across the entire Cygnus-X sample, we find 68.4\% of high luminosity protostars in high stellar density environments.  This shows that the sources with luminosities exceeding 10 $L_{\odot}$ are more likely to be in crowded regions than lower luminosity sources.

We compare the luminosity functions of the protostars in higher stellar density environments with luminosity functions of protostars in lower stellar density environments in Figure \ref{fig:hl_clust_iso}. We see an excess of high luminosity sources in high stellar density regions above $L_{cut}$ in each of the Cygnus-X, CygX-North, and CygX-South luminosity functions.  To determine if this difference is statistically significant, we run the K-S test on the 1000 realizations of each luminosity function and thus retain a distribution of 1000 K-S probabilities. Only nebulosity filtered protostars above $L_{cut}$ are used in the test. The median K-S probabilities are given in Table \ref{table:probs}. In the entire Cygnus-X sample and  the CygX-North sample the probabilities are low enough  to rule out that the high and low stellar density luminosity functions are from the same parent distribution. In CygX-South, the probability that they are from the same distribution is 0.0135, equivalent to a 2.47 $\sigma$ difference in Gaussian statistics.  If we increase the value of $L_{cut}$ to  3~L$_{\odot}$, the resulting probabilities that the high and low density regions change to $log(prob) = -2.7$, $log(prob) = -5.7$, and $prob = 0.10$ for the total region, CygX-North and CygX-South, respectively.   Thus, the difference remains highly significant in the total Cygnus-X region and CygX-North, but the result for CygX-South is no longer significant.

The 24~$\mu$m magnitude distribution for Class II sources  can  be used to test whether spatially varying incompleteness can affect our comparison of high and low stellar density regions.  Figure \ref{fig:hl_m24} shows the $m_{24}$ distributions of protostars in high and low stellar density environments for the entire Cygnus-X region. Based on the K-S test, the probability that the high and low stellar density $m_{24}$ distributions are from the same parent distribution is $log(prob)$ = -3.29; this is similar to the probability returned by the K-S test for the high and low stellar density luminosity functions.   Since the $m_{24}$ distribution of Class II objects, which depend on the masses and ages of the stars and the evolving properties of their disks, are not expected to strongly depend on environment, the Class II objects can be used as a control sample to test for the effects of incompleteness.  Class II sources are identified in Cygnus-X using the same technique as KMG12. This technique selects Class II sources within $D_{c}$ of a protostar candidate and classifies them as high or low stellar density based on the classification of the nearest protostar candidate as residing in a high or low density environment. Since we are unable to apply the equations estimating protostar luminosity to the Class II sources, we do not determine Class II luminosities and therefore we do not impose $L_{cut}$ in this comparison.  Instead we use our 90\% completeness limiting magnitude, $m_{24}$ = 5.07 $mag$, and reject sources with $m_{24}$ fainter than this limit in the comparison.  We perform this comparison of the Class II objects for only the entire Cygnus- X cloud YSO sample since there are only a few Class II sources in the CygX-North and CygX-South regions with $A_{V}$ $>$ 3 and $m_{24}$ $<$ 5.07 $mag$.

A K-S test comparing the distributions of $m_{24}$ of Class II sources with $A_{V} > 3$ within $D_{c}$ of a protostar candidate in a high or low stellar density environment reveals a probability of 0.46 that the Class II high and low stellar density $m_{24}$ distributions are from the same parent distribution. This suggests that incompleteness is not affecting our analysis of the protostars.  However, the number of Class II objects is 48/63 in high/low density regions, respectively; this is much smaller than the 386/245 protostars found in high/low density regions.  To assess the effect of the smaller numbers on the Class II sample,  we run a K-S test between the $m_{24}$ distributions of protostars in high and low stellar density environments, but we randomly select from the full sample of protostars sub-samples for both the high and low density samples equal to the number of Class II in high and low density environments, respectively.  We repeated this 100 times using different randomly chosen protostar candidates;  from these 100 trials the median probability returned by the KS test was 0.13.  Thus, the probability for the reduced protostars sample is smaller than that for the Class II sample, but the K-S test for a smaller sample of protostars would result in a much less significant difference between high and low density region.  In conclusion, the lack of variation of the Class II $m_{24}$ distribution with environment is consistent with our interpretation that the luminosity function of protostars is different between high and low density environments, and not the result of biases in the data.  However, given the smaller sample size for the Class II objects, the use of the Class II objects as a control sample does not provide a definitive statistical test.

Another consideration is whether extinction can play a role.  As noted in KMG12, the extinction does effect the luminosity of individual sources, but a correction for the extinction to a protostar cannot be reliably determined.  However, the extinction is higher in denser, more crowded regions.  Specifically, in our high YSO density regions, 10\% of the sources are projected
on regions with $A_V > 10$, 3.2\% on $A_V > 20$ and 0.8\% on $A_V > 30$.  In comparison, for the low density regions,
6.5\% of the protostars are projected on regions of $A_V > 10$, 1.3\% on regions of $A_V > 20$ and none have $A_V > 30$.  Since crowded regions  have higher extinction they should contain sources with systematically underestimated luminosities. Consequently,  extinction cannot explain the   higher protostellar luminosities in the dense environments. 

We conclude that in the total Cygnus-X sample and the CygX-North region there are significant differences in the 2-1000~L$_{\odot}$  luminosity functions between regions of high and low stellar density, with the luminosity function in regions of high stellar density biased to higher luminosities. In the CygX-South region, the evidence for such a change is inconclusive.  We note that some the highest luminosity sources are saturated and are not included in this work; thus, we cannot address the possibility of variations in the very high end of the luminosity function  above 1000~L$_{\odot}$. 

\subsection{Comparison With Star-Forming Clouds Within 1 Kpc}

How does the protostellar luminosity functions of the Cygnus-X star formation complex compare with the less massive clouds within 1 kpc of our Sun?  KMG12 determined the luminosity functions of protostars toward 9 molecular clouds out to a distance of 830 pc.  Their sample consists of clouds that formed both high and low mass stars (Orion, Cep OB3, and Mon R2, hereafter: HMSFC) as well as clouds containing only low mass protostars (Serpens, Perseus, Ophiuchus, Lupus, Chamaeleon, and Taurus, hereafter: LMSFC).  KMG12 combined these luminosity functions into two composite luminosity functions, one for the clouds with high mass stars and one for the clouds without high mass stars.  Since these were made using a technique similar to that used to determine the Cygnus-X protostellar luminosity function, these two luminosity functions provide the means to compare Cygnus-X to local star formation.

First we compare the shape of the luminosity functions.  The luminosity functions for Cygnus-X, CygX-North, CygX-South, and those for the LMSFC, HMSFC, and Orion from KMG12 are shown in Figure \ref{fig:lum_frac}.  The combined luminosity function of HMSFCs from KMG12 rises toward our completeness limit for Cygnus-X, $L_{cut}$ = 2.16 $L_{\odot}$, and has a tail extending toward luminosities above 100 $L_{\odot}$.  The LMSFC luminosity function also rises toward our Cygnus-X completeness limit, $L_{cut}$ = 2.16 $L_{\odot}$, but does not show a tail at higher luminosities.  We see that the Cygnus-X, CygX-North, and CygX-South protostellar luminosity functions rise toward lower luminosity as well.  At the high luminosity end of the distributions, we see tails extending toward luminosities above 1000 $L_{\odot}$, most noticeably in CygX-North.

For each of the nine regions in KMG12 we also have 1000 realizations of the protostellar luminosity functions.  Each of the 1000 realizations are used because of the uncertainty in the reddened disk contamination removal.  We use only the luminosities above $L_{cut}$, above which neither the HMSFC luminosity function nor the LMSFC luminosity function show a peak.   It should be noted that the LMSFC luminosity function has only 32 protostars with luminosities above $L_{cut}$ and does not contain protostars with luminosity above 100 $L_{\odot}$.  The HMSFC luminosity function contains 150 protostars above $L_{cut}$, 11 of which are above 100 $L_{\odot}$.  KMG12 found that a K-S test suggests a real difference between the high mass SF cloud and low mass SF cloud luminosity functions when using a luminosity cutoff less than 0.05 $L_{\odot}$; however, it should be noted that above $L_{cut}$ = 2.16 $L_{\odot}$ the HMSFC and LMSFC luminosity functions give a median probability of 0.8971 of coming from the same parent distribution. 

The results of K-S tests comparing the luminosity functions of the Cygnus-X, CygX-North, and CygX-South protostar samples with the HMSFC and LMSFC luminosity functions are given in Table \ref{table:probs}.  The Cygnus-X luminosity function has a higher median probability when compared to the HMSFC luminosity function (0.6573) than when compared to the LMSFC luminosity function (0.0360).   In CygX-North, we find lower probabilities ($<<$ 0.01) when comparing the protostar luminosity function with those of the HMSFC and LMSFC, and can rule out the possibility that the luminosity function from CygX-North is from the same parent distribution as either the HMSFC or LMSFC luminosity function.  In CygX-South there is a low probability (0.0111) that the CygX-South luminosity function is from the same parent distribution as the LMSFC luminosity function, but a fairly large probability (0.8123) that it is from the same parent distribution as the HMSFC luminosity function.

 Neither the Cygnus-X nor the CygX-South protostellar luminosity function has a high probability of being from the same parent distribution as the LMSFC protostellar luminosity function of KMG12, though there is a greater than 60\% probability that both are like the HMSFC luminosity function.    The CygX-North protostellar luminosity function shows a very low probability of being from the same parent distribution as any of the other protostar samples given in this work or KMG12.  

\section{Discussion on Protostar Luminosity and Environment}
The fundamental question addressed by this paper is whether the properties of protostars are influenced by their environment.  This was motivated by the work of KMG12, who compared luminosity functions of protostars in nearby ($<$1 kpc) star-forming clouds in different environments and found evidence that the luminosity function did depend in environment (see Section \ref{sec:intro}).  In this work, we sought to determine whether the luminosity functions of protostars in Cygnus-X and within sub-regions of Cygnus-X vary across the cloud complex. This study identifies and utilizes the largest sample of protostars within 1.5 kpc that are found at a common distance. It not only includes a larger protostar candidate population than KMG12, but also studies a cloud complex with more extreme star formation conditions than found in the KMG12 sample. 

Within the Cygnus-X complex, the star-forming regions show distinct morphologies suggestive of different star-forming environments.  In particular, the CygX-North and CygX-South regions exhibit different distributions of gas, stars and mid-IR nebulosity.  The most luminous protostar candidates (L $>$ 10 $L_{\odot}$) in CygX-South are concentrated near the edges of the molecular cloud closest to Cyg OB2 where the bright mid-IR nebulosity shows that the cloud is directly interacting with the Cyg OB2 association; these may be sites of triggered star formation.  In comparison, the most luminous protostar candidates in CygX-North are found clustered in filamentary cloud structures located throughout the region, and are often in regions which do not show the bright nebulosity that would be expected from cloud surfaces directly illuminated by the OB association. \citet{2007A&A...476.1243M} also noted a difference between the CygX-North and Cyg-South regions.  They found that a higher fraction of the mass in the CygX-North region was concentrated into cores and dense structures than in Cyg-South; suggesting that the formation of cores was 5 times more efficient in CygX-North.  The lack of dense gas may explain why much of the CygX-South cloud is comparatively quiescent, with the star formation concentrated on the edges of the clouds where potentially dense gas can be created by the compression of the cloud surfaces. The CygX-North regions, in contrast,  contains a dense, massive ridge of gas which appears to accreting gas from surrounding filamentary structures  \citep{2010A&A...520A..49S,2012A&A...543L...3H}. The distinctive morphologies of these regions suggests that both triggered and spontaneous modes of star formation are operating in Cygnus-X. 

We observe not only morphological differences between these two regions (Figures \ref{fig:av_map}, \ref{fig:CO_map}, \ref{fig:north_map}, and \ref{fig:south_map}), but also statistically significant differences in the luminosity functions of protostars in these two regions (see Section \ref{sec:n_s}). These differences in the luminosity functions of protostars in dissimilar environments suggest that the properties of the protostars do indeed depend on their environment. 

We also examine the effect of stellar density on protostellar luminosity by comparing the samples of protostars found in high and low stellar density regions.  Again, we find statistically significant differences in the luminosity functions of protostars as a function of their environment.  As found in KMG12 for in the Orion molecular clouds, the luminosity function in regions of high stellar density is biased  to higher luminosities in the Cygnus-X and CygX-North samples.  Thus, two separate analyses, the comparison of the CygX-South and CygX-North regions and the comparison of regions of high and low stellar density, demonstrate that there is a dependence of protostellar luminosity on environment in Cygnus-X. 

{ Compared to the molecular clouds within 1 kpc of the Sun studied by KMG12, we find that none of the luminosity functions found in Cygnus-X are similar to the luminosity functions of the $< 1$~kpc molecular clouds without massive stars.  The Cygnus-X and CygX-South are similar to the luminosity functions of $< 1$~kpc clouds with massive stars, this result reinforces the dichotomy between low mass and high mass star-forming clouds found by KMG12.  Furthermore, the CygX-North luminosity function is different from the other Cygnus-X regions and all $< 1$~kpc clouds; suggesting that the conditions for star formation in CygX-North must be different than these other regions to create such a different population of protostars.}

Do these differences imply differences in the masses of the emerging stars? Since protostellar luminosity is a combination of intrinsic stellar luminosity and accretion luminosity, there is not a direct relationship between luminosity and either the instantaneous or the final mass of the protostar.  Although the intrinsic luminosity will increase with mass \citep[e.g.][]{1997ApJ...475..770H,1993ApJ...418..414P}, the accretion luminosity depends on both the accretion rate and the mass of the star.  Hence a high accretion luminosity implies a higher stellar mass or a higher accretion rate.  Given that the final mass of a protostar depends on the accretion rate and the duration of the accretion, a higher accretion rate does not necessarily lead to a more massive star. Instead, differences in the observed luminosity functions may be due to differences in ultimate masses of the stars, or they may simply result from differences in the duration of the protostellar accretion phase.
 
Do current models predict that we should see more luminous protostars primarily in high stellar density regions?  KMG12 compared the HMSFC luminosity function, which is similar in shape to the Orion luminosity function, with turbulent core, two-component turbulent core, and competitive accretion models from \cite{2011ApJ...736...53O}.  They found that the competitive accretion model fit the HMSFC luminosity function best. While the competitive accretion model did not predict well the height of the peak of the HMSFC luminosity function, it did predict the location of the peak and width of the luminosity function. The competitive accretion model \citep{2004MNRAS.349..735B} also  predicts that the most massive protostars (and hence the most luminous protostars) are found toward the dense centers of clusters.  Alternatively, \cite{2011ApJ...739...84G} found that regions with a high surface density of YSOs have higher gas column densities and higher star formation efficiencies.  Hence, the high stellar density regions in Cygnus-X likely contain higher column density of molecular gas; these higher column densities in turn would imply higher infall rates and masses \citep[see][]{2010ApJ...713.1120K,2003ApJ...585..850M}.  Regardless, of the specific models, these observations suggest a scenario where protostars in regions of increasingly higher stellar density and gas density can have higher accretion rates, potentially leading to the formation of more massive stars.  The high accretion rates may in turn be fed by gas flowing through sub-filaments feeding into highly active star-forming regions, as observed in the DR21 filament by \citet{2010A&A...520A..49S}.  They also suggests that searches for variations on the IMF should concentrate on comparing high and low stellar density regions.  Interestingly, both simulations and observations now suggest that the IMF may be different for the large number of stars forming outside of large clusters in more diffuse, less strongly gravitationally bound, star-forming regions \citep{2008MNRAS.386....3C,2011MNRAS.410.2339B,2012ApJ...752...59H,2013ApJ...764..114H}.

\section{Conclusions}

\begin{figure}
\epsscale{1.0}
\plotone{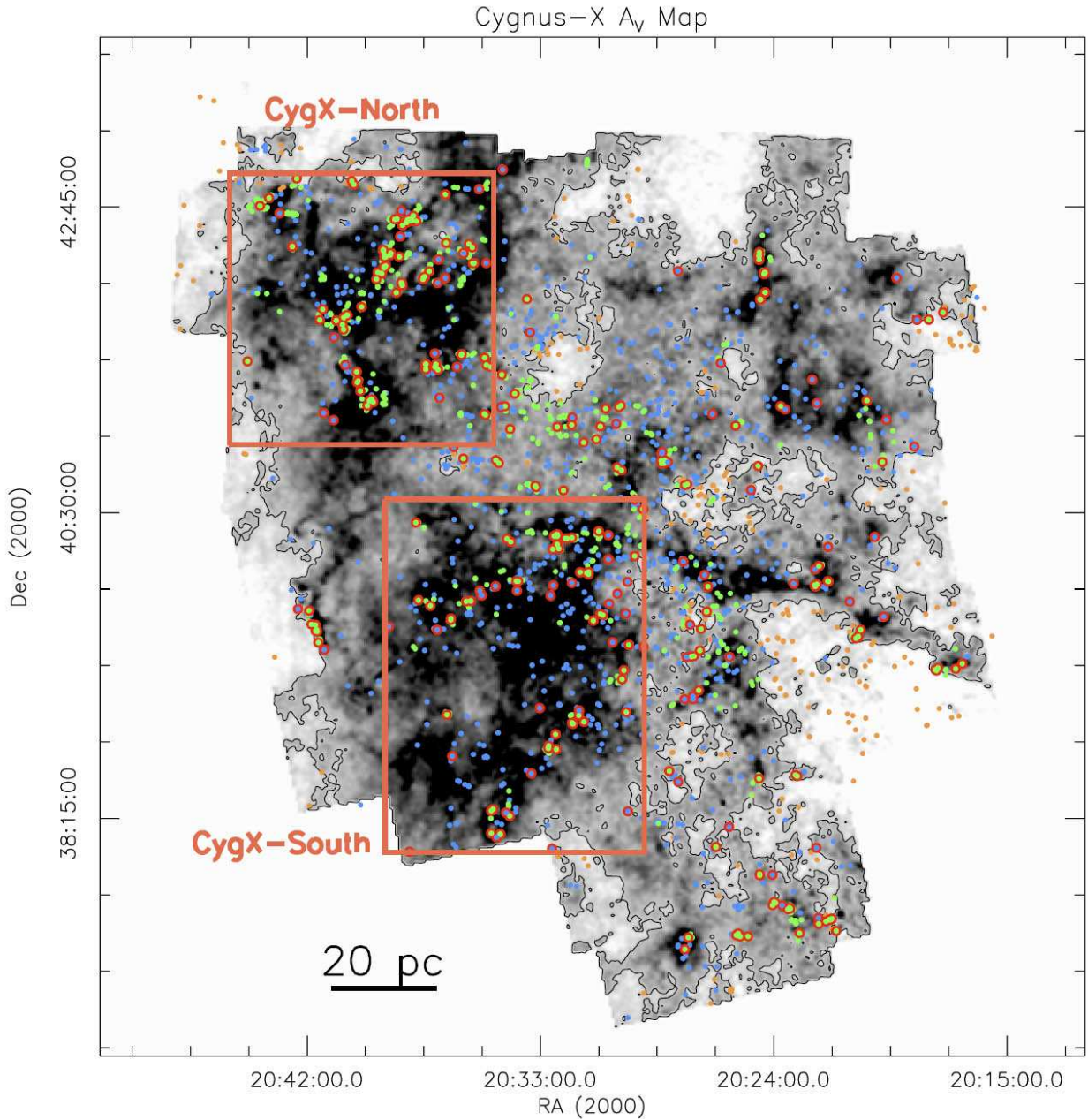}
\caption{Cygnus-X $A_{V}$ map with IRAC and MIPS coverage area.  The $A_V$ map is rendered in a greyscale with the $A_V = 3$ contour  displayed.   Shown are protostar candidates  towards regions with $A_{V} > 3$ and in either high stellar density environments (green dots) or low stellar density environments (blue dots).  Protostar candidates projected on regions with $A_{V} < 3$ (orange dots) are not used in this analysis. Protostar candidates with luminosity $>$ 10 $L_{\odot}$ are marked in red circles.  The CygX-North and CygX-South sub-regions are bordered in the salmon boxes.}
\label{fig:av_map}
\end{figure}

\begin{figure}
\epsscale{1.0}
\plotone{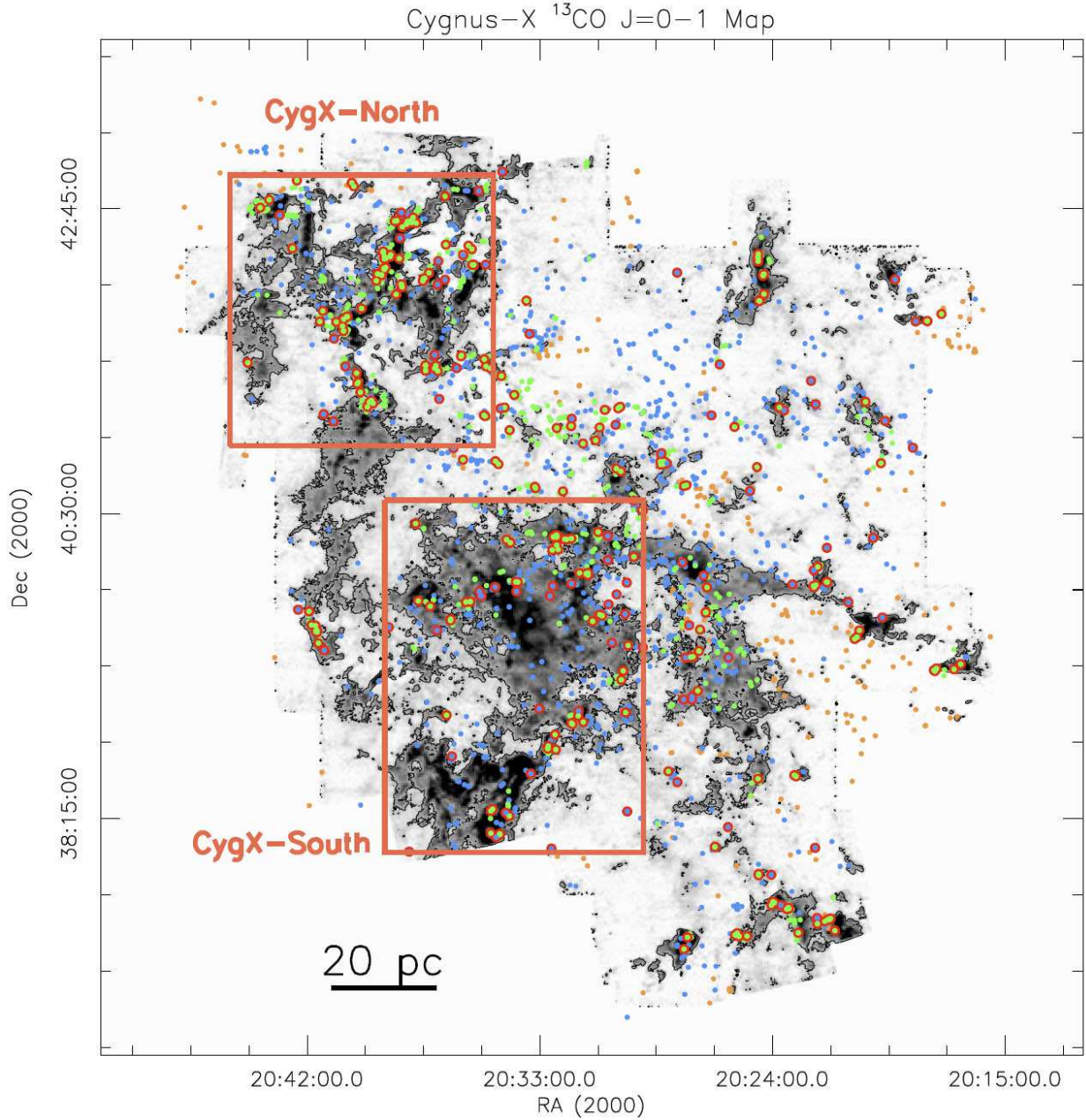}
\caption{Cygnus-X $^{13}$CO J = 1$\rightarrow$0 {\it velocity integrated intensity map} from \cite{2011A&A...529A...1S} with IRAC and MIPS coverage area. The markers and boxes are the same as in Figure 3.}
\label{fig:CO_map}
\end{figure}

\begin{figure}
\epsscale{1.0}
\plotone{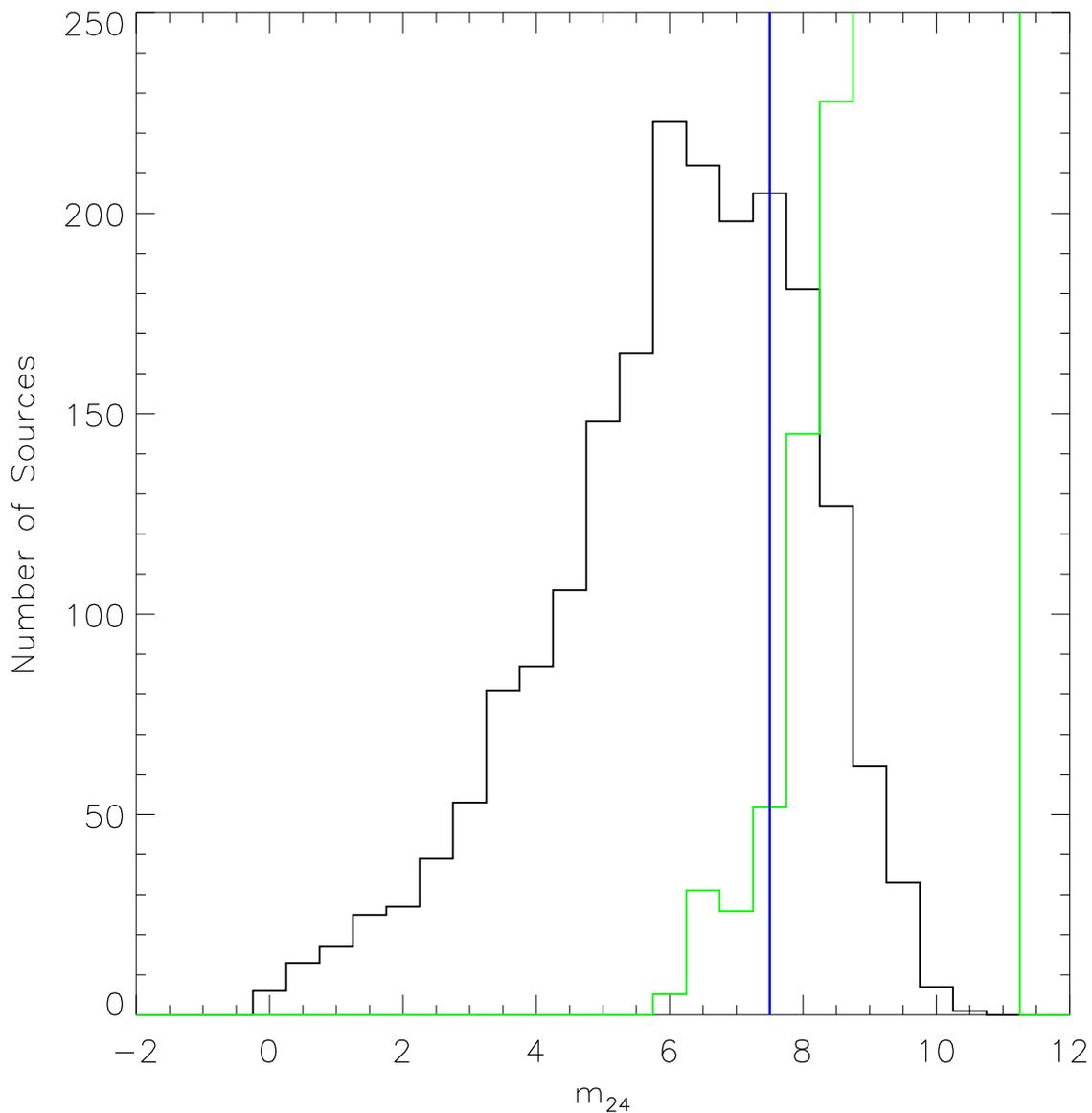}
\caption{The distributions of $m_{24}$ for all sources which fit the protostar selection criteria and which are in regions with $A_{V} > 3$ (black) and for SWIRE galaxy sources fitting the protostar criteria (green).  The SWIRE distribution is scaled so to represent the distribution of galaxies for a field equivalent in size to the Cygnus-X survey.  The vertical lines shows our choice of the  $m_{24}$ cutoff (vertical blue bar), $m_{24}$ = 7.5.}
\label{fig:24_cutoff}
\end{figure}

\begin{figure}
\plotone{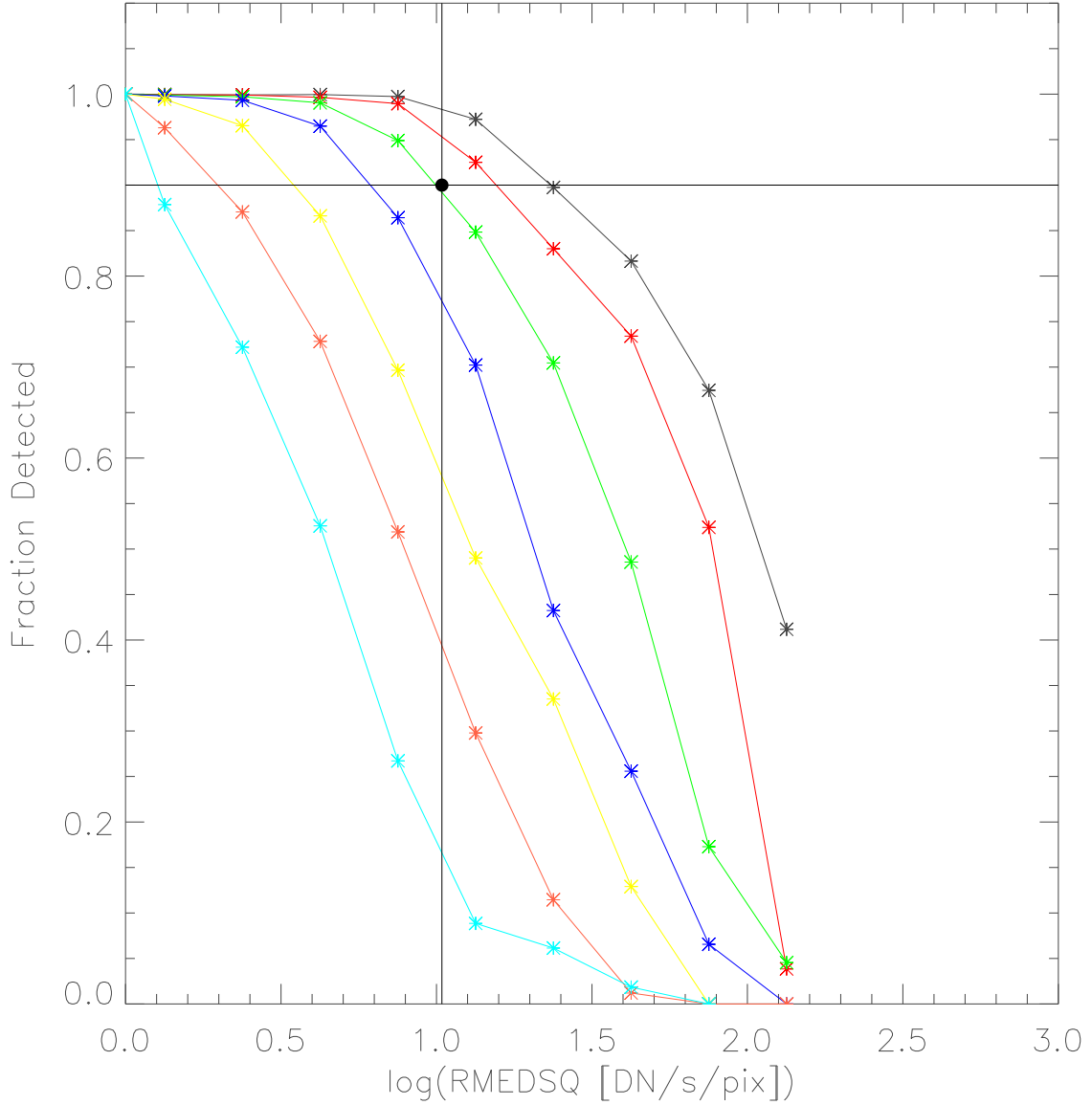}
\caption{Fraction of fake stars recover at each test magnitude vs. log(RMEDSQ). The curves are for magnitude (from the leftmost curve) $m_{24}$ = 7.125, 6.625, 6.125, 5.125, 4.625, 4.125.  The $m_{24}$ = 3.125 and $m_{24}$ = 3.325 test magnitudes did not show a decrease at high RMEDSQ and are not shown.  The measured points are given by the asterisks and the lines are linear interpolations between those points.  The horizontal bar shows 90\% completeness and vertical bar shows RMEDSQ = 10.41.  The $m_{24}$ cutoff corresponding to RMEDSQ = 10.41, $m_{24}$ = 5.07, is shown as a filled circle.  We find that the fraction of sources detected declines with increasing RMEDSQ.}
\label{fig:MAD_mag}
\end{figure}

\begin{figure}
\plotone{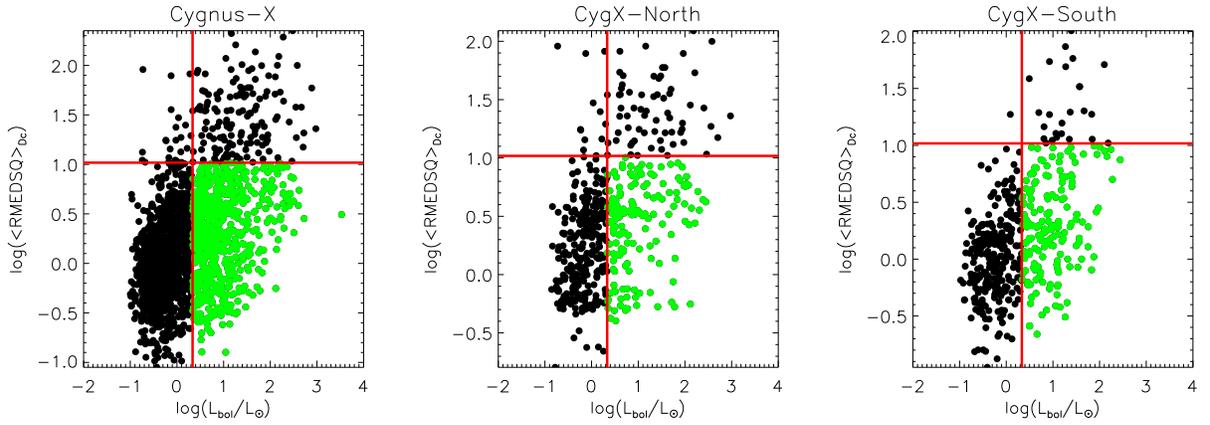}
\caption{Log($<$RMEDSQ$>_{D_{c}}$) vs. log($L_{bol}$/L$_{\odot}$) for protostar candidates in the full Cygnus-X sample (left), CygX-North sample (middle), and CygX-South sample (right).  All protostar candidates projected on regions with $A_{V} > 3$ are shown in black or green. The luminosity cutoff $L_{cut}$ and corresponding $<$RMEDSQ$>_{D_{c}}$ cutoff are shown by the red lines.  The sources below the $<$RMEDSQ$>_{D_{c}}$ decrease are the nebulosity filtered sample.  The nebulosity filtered sources which are above $L_{cut}$ are used in our comparison of the luminosity functions; these are colored in green.}
\label{fig:rmedsq}
\end{figure}

\begin{figure}
\plotone{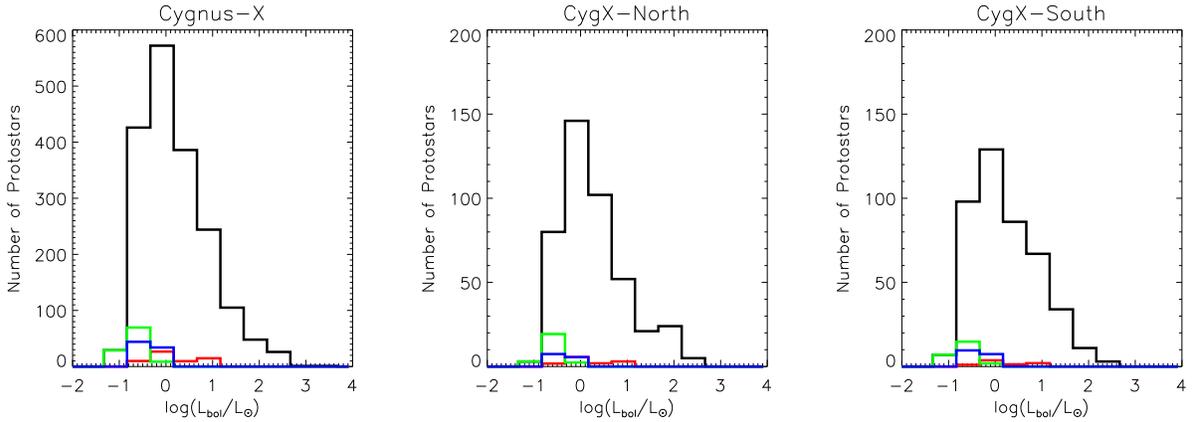}
\caption{Luminosity functions for nebulosity filtered protostar candidates with contamination in the full Cygnus-X sample (left), CygX-North sample (middle), and CygX-South sample (right).  Protostellar luminosity functions for the protostellar candidates which contain some contamination, are shown in black.  The color histograms show the estimated luminosity functions of contamination from reddened Class II contamination (red), edge-on Class II sources (green), and star-forming galaxies (blue).}
\label{fig:blum_cont}
\end{figure}

\begin{figure}
\plotone{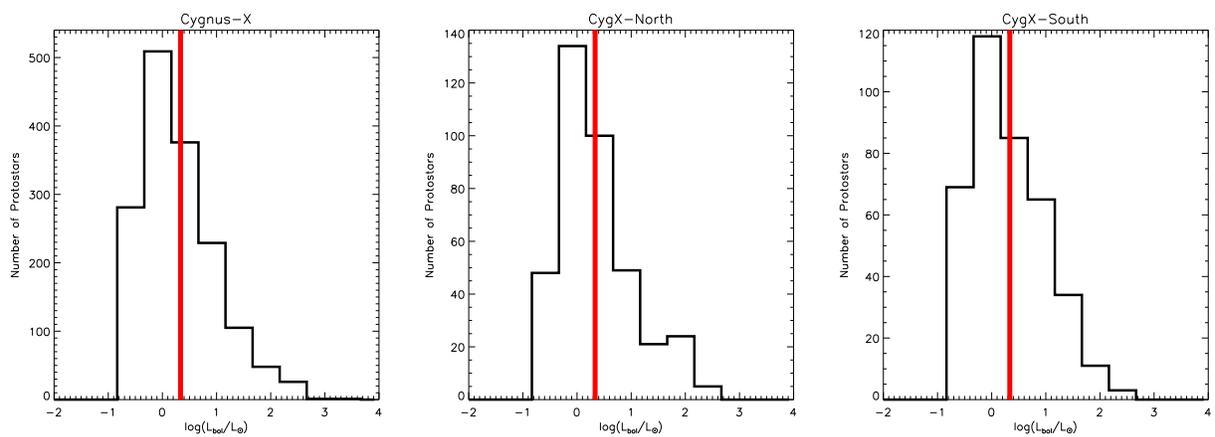}
\caption{Luminosity functions of nebulosity filtered protostars with estimated luminosities of contamination from reddened disk sources, edge-on Class II sources, and background galaxies removed for the Cygnus-X sample (left), CygX-North sample (middle), and CygX-South sample (right).  The vertical line shows the limiting $L_{bol}$ ($L_{cut}$) found as the 90\% completeness limit in Section \ref{sec:RMEDSQ}.}
\label{fig:blum_sub_hists}
\end{figure}

\begin{figure}
\epsscale{1.0}
\plotone{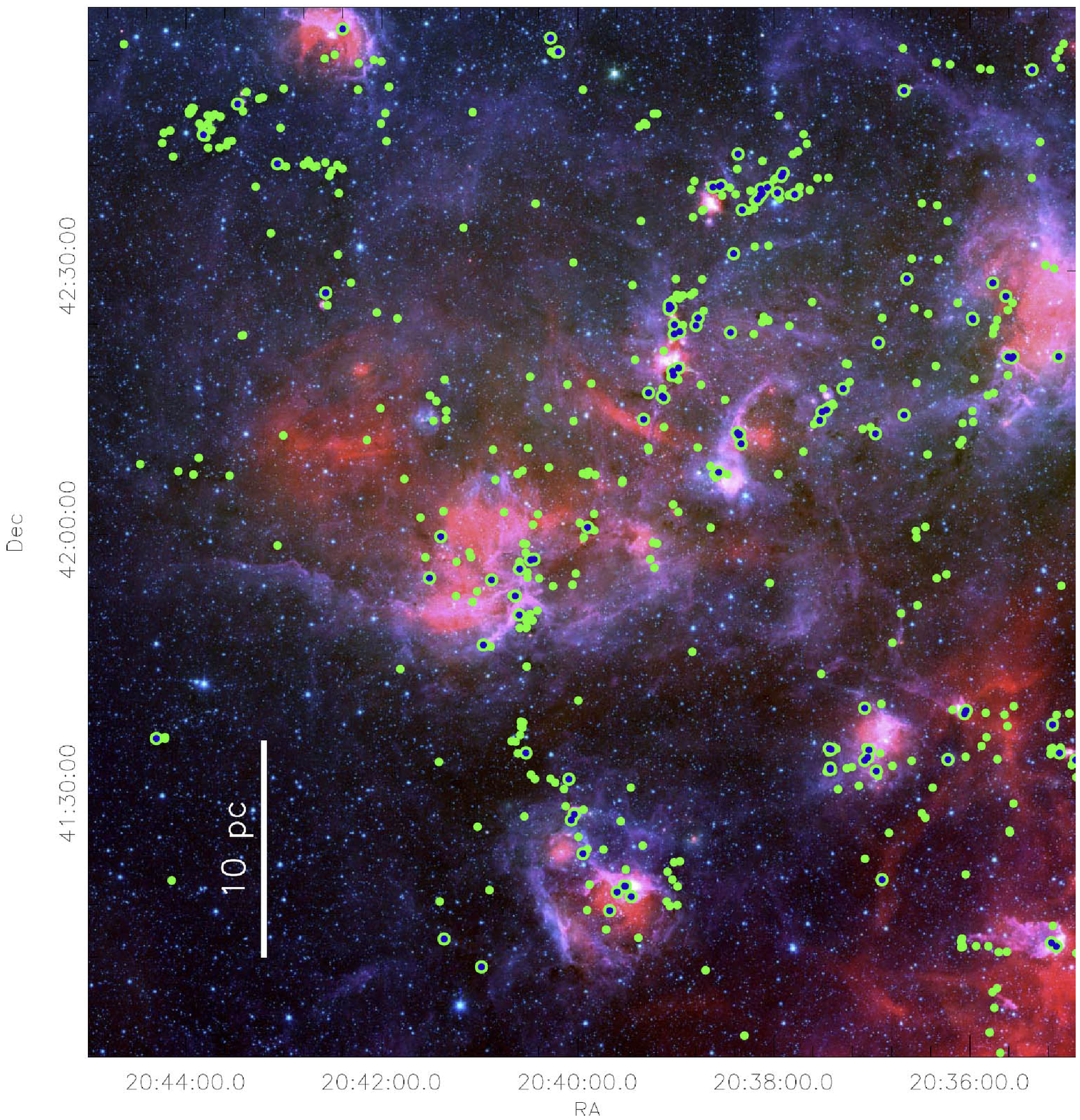}
\caption{{\it Spitzer} 3.6 $\mu$m (blue), 4.5 $\mu$m (green), and 24 $\mu$m (red) image of CygX-North.  Protostar candidates projected on regions with $A_{V} > 3$ are shown as green filled circles; those with luminosities above 10 $L_{\odot}$ are shown with blue centers.}
\label{fig:north_map}
\end{figure}

\begin{figure}
\epsscale{1.0}
\plotone{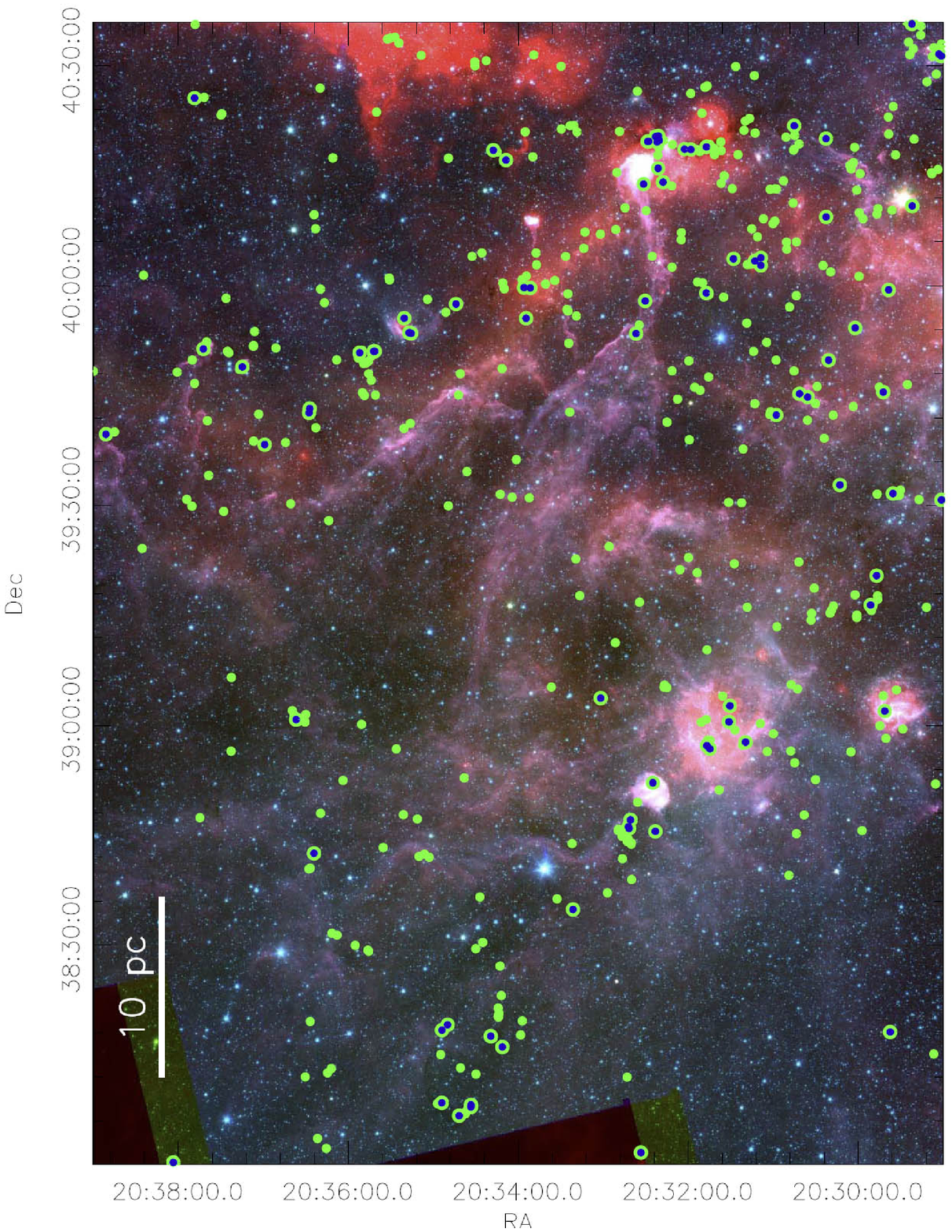}
\caption{{\it Spitzer} 3.6 $\mu$m (blue), 4.5 $\mu$m (green), and 24 $\mu$m (red) image of CygX-South.    Protostar candidates projected on regions with $A_{V} >  3$ are shown as green filled circles; those with luminosities above 10 $L_{\odot}$ are shown with blue centers.}
\label{fig:south_map}
\end{figure}

\begin{figure}
\epsscale{1.0}
\plotone{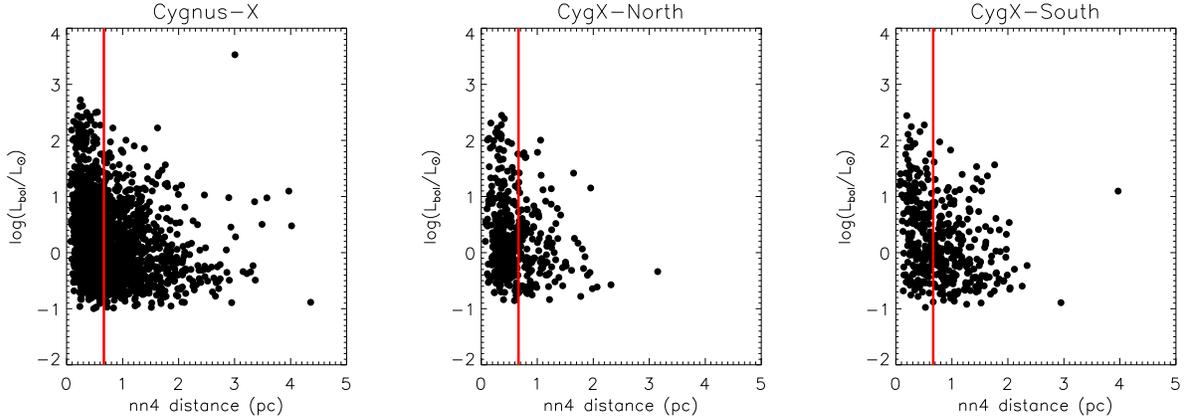}
\caption{Nearest neighbor distances (for 4$^{th}$ nearest-neighbor YSO) and calculated bolometric luminosity for the nebulosity filtered protostar candidates ($<$RMEDSQ$>_{D_{c}}  < 10.41$) in the entire Cygnus-X sample (left), CygX-North sample (middle), and CygX-South sample (right).  The median protostar nn4 distance $D_{c}$ is shown as a vertical red line, separating the sample into high and low stellar density populations.}
\label{fig:nns_all}
\end{figure}

\begin{figure}
\epsscale{1.0}
\plotone{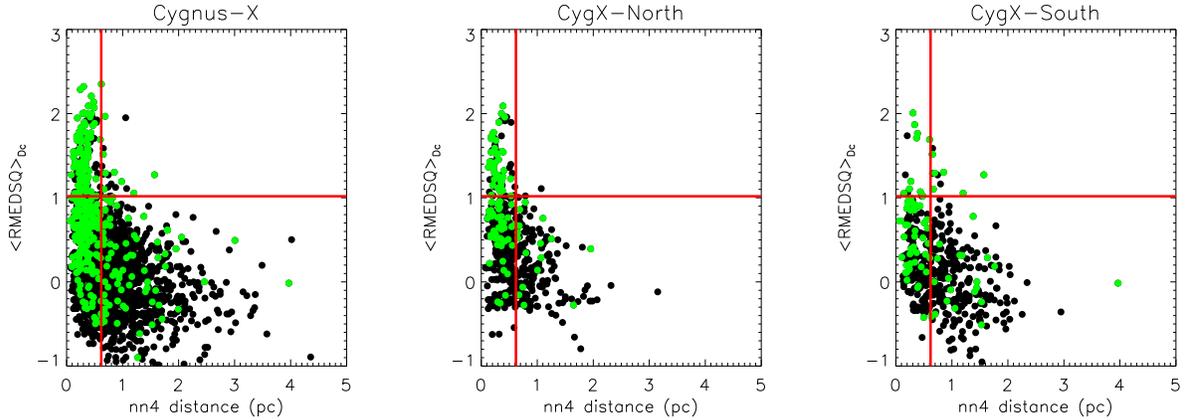}
\caption{$<$RMEDSQ$>_{D_{c}}$ vs. nn4 distance for the entire Cygnus-X protostar candidate sample (left), CygX-North protostar candidate sample (middle), and CygX-South protostar candidate sample (right).  The horizontal bar indicates the location of our RMEDSQ cut above which we reject sources due to nebulosity discussed in Section \ref{sec:RMEDSQ}.  The vertical bar shows $D_{c}$, the division between high and low stellar density populations.  The protostar candidates with luminosity greater than 10 $L_{\odot}$ are shown in green.  We find the sources with highest $<$RMEDSQ$>_{D_{c}}$, regardless of luminosity, in regions of higher stellar density.}
\label{fig:MAD_nns}
\end{figure}

\begin{figure}
\plotone{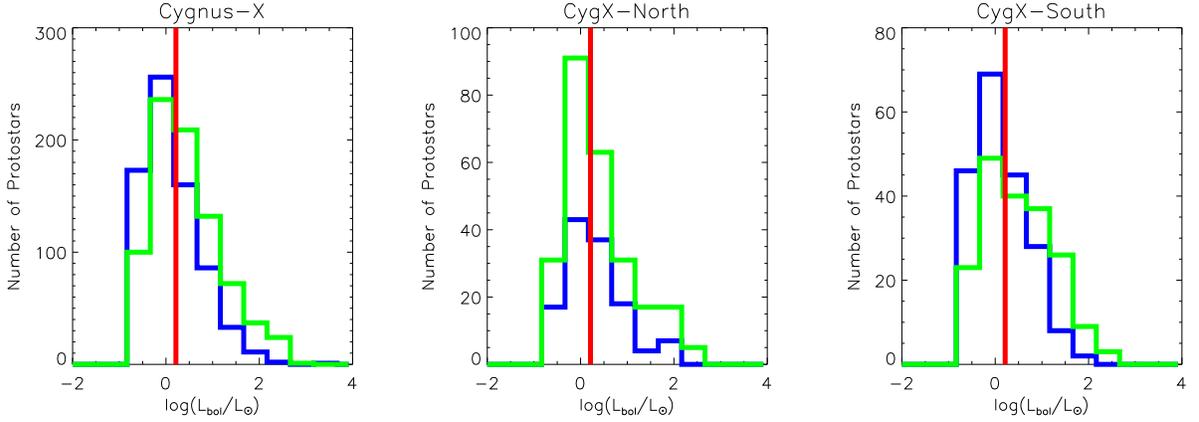}
\caption{Contamination subtracted luminosity functions of the nebulosity filtered protostars ($<$RMEDSQ$>_{D_{c}}$ less than 10.41) in high (green) and low (blue) stellar density environments in the entire Cygnus-X cloud sample (left), CygX-North sample (middle), and CygX-South (right).  The vertical line is $L_{cut}$, the 90\% completeness limit.  Each distribution is the average of the 1000 Monte Carlo trials of the contamination subtraction.}
\label{fig:hl_clust_iso}
\end{figure}

\begin{figure}
\epsscale{1.0}
\plotone{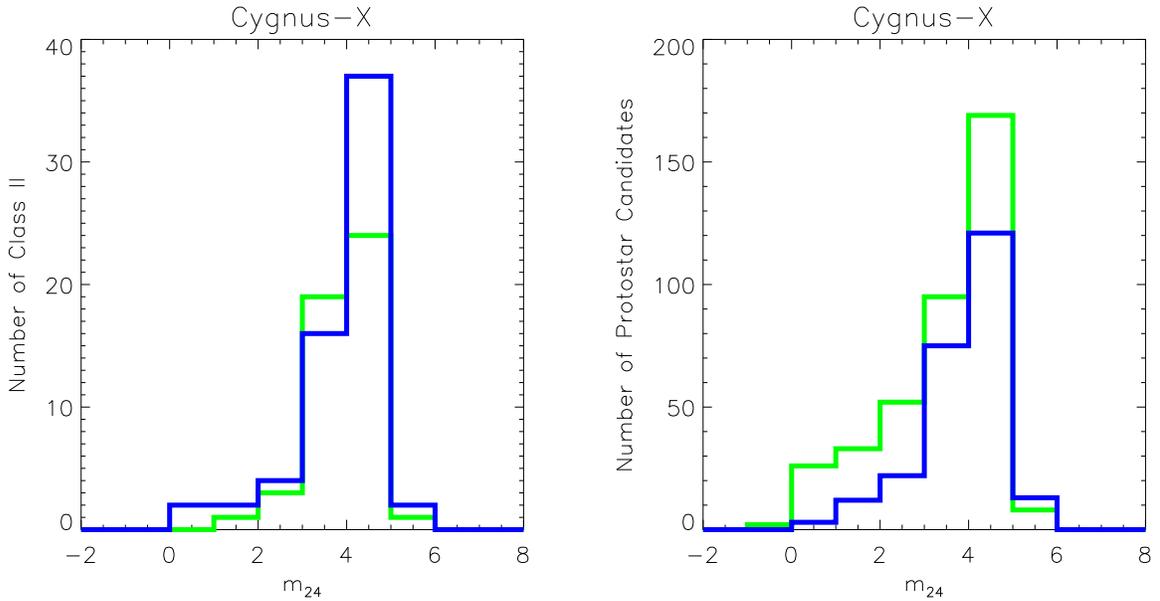}
\caption{ The distributions of $m_{24}$ for YSOs in Cygnus-X.  {\bf Left:} Distributions of $m_{24}$ for Class II sources within 0.62 $pc$ ($D_{c}$) of  nebulosity filtered protostars with $A_{V} > 3$ in high stellar density (green) or low stellar density (blue) environments.  {\bf Right:} the distributions of $m_{24}$ for nebulosity filtered protostars with $A_{V} > 3$ in high stellar density (green) and low stellar density (blue) environments.}
\label{fig:hl_m24}
\end{figure}

\begin{figure}
\epsscale{1.0}
\plotone{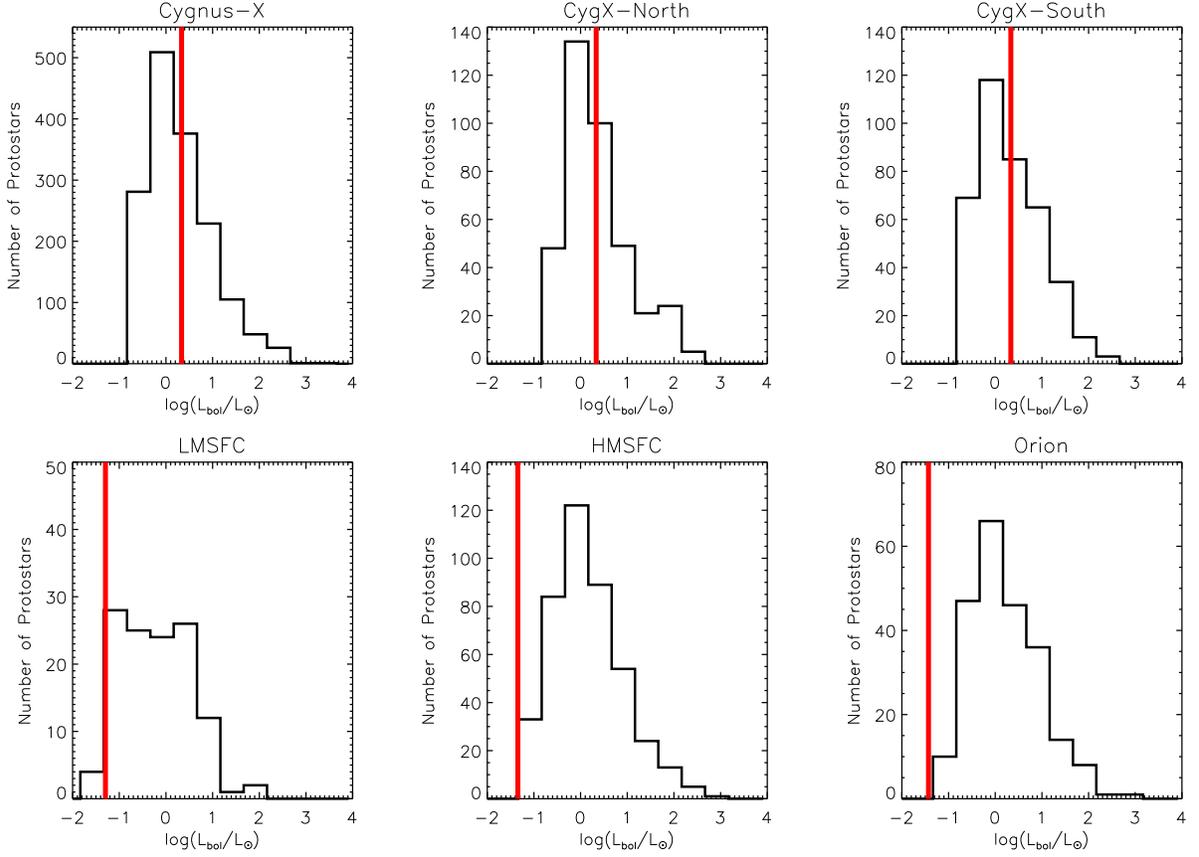}
\caption{Luminosity functions for Cygnus-X, CygX-North, CygX-South (top row),  low mass star-forming clouds within 1~kpc of the Sun (LMSFC), high mass star-forming cloud within 1~kpc of the Sun (HMSFC), and the Orion molecular clouds (bottom row).  The HMSFC, LMSFC, and Orion luminosity functions are from KMG12.  The red bars show the completeness limit in each region.}
\label{fig:lum_frac}
\end{figure}

We identify over 1800 protostar candidates in the Cygnus-X star-forming cloud complex.  These candidates, as shown in Figures \ref{fig:av_map} and \ref{fig:CO_map}, are distributed throughout the complex of molecular clouds observed towards Cygnus-X.  We estimate luminosities for these protostar candidates and create luminosity functions corrected for contamination due to reddened Class II sources, edge-on disk sources, and galaxies.  We present contamination-subtracted luminosity functions for the entire Cygnus-X cloud complex and for the protostars found within either the CygX-North or CygX-South molecular clouds.  Using the median nn4 distance, the distance from any protostar candidate to the 4$^{th}$ nearest-neighbor Class II or protostar candidate, we designate half our protostar candidate sample as residing in a ``high stellar density environment'' and half as residing in a ``low stellar density environment'' (Figures~\ref{fig:av_map} and \ref{fig:CO_map}).  Contamination subtracted luminosity functions are then constructed for the low and high density regions.  Our results are listed below.

\begin{itemize}
\item{We find that the entire Cygnus-X protostar sample luminosity function increases toward lower luminosity down past our completeness limit near 2 $L_{\odot}$, and exhibits a tail leading toward luminosities above 1000 $L_{\odot}$.  The CygX-North and CygX-South luminosity functions also show these general characteristics, though the CygX-South rises less steeply toward lower luminosity.}

\item{The spatial distribution of protostars in CygX-South and CygX-North exhibit distinct differences.  The CygX-South cloud shows many of its protostar candidates at the edges of the cloud, particularly the protostars with luminosities above 10 $L_{\odot}$, while the CygX-North cloud shows the protostars concentrated in large filamentary structures.}

\item{We compare the Cygnus-X, CygX-North, and CygX-South protostar luminosity functions using a Kolmogorov-Smirnov (K-S) test.  We can rule out the possibility that the CygX-North protostar luminosity function is from the same parent distribution as the CygX-South or full Cygnus-X protostar sample luminosity function.   We are not able to rule out the possibility that the Cygnus-X and CygX-South protostar luminosity functions are from the same parent distribution.}

\item{We also compare the Cygnus-X, CygX-North, and CygX-South protostar luminosity functions to the luminosity functions of protostars in nearby ($<$ 1 kpc) molecular clouds (KMG12).  The K-S tests show a low probability that the Cygnus-X or CygX-South luminosity function comes from the same parent distribution as either the low mass star-forming cloud (LMSFC) luminosity function, but these tests cannot statistically distinguish Cygnus-X or CygX-South luminosity functions from the high mass star-forming cloud (HMSFC) luminosity function.  A K-S test comparing the CygX-North luminosity function with the HMSFC and LMSFC luminosity functions rules out the possibility that they share the same parent distribution.}

\item{Finally, we perform a K-S test comparing the luminosity functions of protostars in high stellar density environments with luminosity functions of protostars in low stellar density environments for each Cygnus-X, CygX-North, and CygX-South protostar samples.  In the case of CygX-North and Cygnus-X, we are able to conclude that it is very unlikely that these two luminosity functions are from the same parent distribution; for  CygX-South the results are inconclusive. In agreement with what was found in the Orion molecular clouds by KMG12, the Cygnus-X and CygX-North protostars found in regions of high stellar densities are biased to higher luminosities.  This is further evidence that the properties of protostars depend on environment, and in particular on the column density of stars and/or gas.}

\item{Our study of the protostellar luminosity function supports one of the primary results of KMG12, that the luminosity function of protostars varies within molecular complexes and is dependent on the local environment in which protostars are found.  These results motivate comparative studies of ensembles of protostars in different environments with the goal of understanding how environmental factors influence the infall and accretion of gas onto protostars and the eventual initial mass function of the protostars.}

\end{itemize}

These observations set the stage for future analyses combining PACS and SPIRE photometry from Herschel surveys of the Cygnus-X region.  By extending the photometry into the far-IR, the Herschel data promise more robust luminosities for the Cygnus-X protostars, detect protostars not identified by Spitzer as well as reliably producing maps of the dense molecular gas from which they are forming \citep[e.g.][]{2010A&A...518L.122F,2012A&A...543L...3H,2012A&A...546A..74D,2013ApJ...767...36S}.

This work is based in part on observations made with the {\it Spitzer Space Telescope}, which is operated by the Jet Propulsion Laboratory, California Institute of Technology under a contract with NASA. Support for the work of STM, EK, RAG, JH and KK was provided by NASA through an award issued by JPL/Caltech. HAS acknowledges partial support for this work from NASA grant   NNX12AI55G.
This publication makes use of data products from the Two Micron All Sky Survey, which is a joint project of the University of Massachusetts and the Infrared Processing and Analysis Center/California Institute of Technology, funded by the National Aeronautics and Space Administration and the National Science Foundation. N.S. and S.B. acknowledge support by the ANR-11-BS56-010 project ``STARFICH''

\clearpage

\clearpage

\clearpage

\clearpage

\begin{thebibliography}

\bibitem[Allen et al.(2004)]{2004ApJS..154..363A} Allen, L.~E., Calvet, N., D'Alessio, P., et al.\ 2004, \apjs, 154, 363 
\bibitem[Beerer et al.(2010)]{2010ApJ...720..679B} Beerer, I.~M., Koenig, X.~P., Hora, J.~L., et al.\ 2010, \apj, 720, 679 
\bibitem[Bressert et al.(2010)]{2010MNRAS.409L..54B} Bressert, E., Bastian, N., Gutermuth, R., et al.\ 2010, \mnras, 409, L54 
\bibitem[Bonnell et al.(2004)]{2004MNRAS.349..735B} Bonnell, I.~A., Vine, S.~G., \& Bate, M.~R.\ 2004, \mnras, 349, 735 
\bibitem[Bonnell et al.(2011)]{2011MNRAS.410.2339B} Bonnell, I.~A., Smith, R.~J., Clark, P.~C., \& Bate, M.~R.\ 2011, \mnras, 410, 2339 
\bibitem[Bontemps et al.(2001)]{2001A&A...372..173B} Bontemps, S., Andr{\'e}, P., Kaas, A.~A., et al.\ 2001, \aap, 372, 173 
\bibitem[Bontemps et al.(2010)]{2010A&A...524A..18B} Bontemps, S., Motte, F., Csengeri, T., \& Schneider, N.\ 2010, \aap, 524, A18
\bibitem[Calvet et al.(1994)]{1994ApJ...434..330C} Calvet, N., Hartmann, L., Kenyon, S.~J., \& Whitney, B.~A.\ 1994, \apj, 434, 330 
\bibitem[Carpenter(2000)]{2000AJ....120.3139C} Carpenter, J.~M.\ 2000, \aj, 120, 3139 
\bibitem[Chapman \& Mundy(2009)]{2009ApJ...699.1866C} Chapman, N.~L., \& Mundy, L.~G.\ 2009, \apj, 699, 1866 
\bibitem[Clark et al.(2008)]{2008MNRAS.386....3C} Clark, P.~C., Bonnell, I.~A., \& Klessen, R.~S.\ 2008, \mnras, 386, 3 
\bibitem[Comer{\'o}n \& Pasquali(2012)]{2012A&A...543A.101C} Comer{\'o}n, F., \& Pasquali, A.\ 2012, \aap, 543, A101 
\bibitem[Crapsi et al.(2008)]{2008A&A...486..245C} Crapsi, A., van Dishoeck, E.~F., Hogerheijde, M.~R., Pontoppidan, K.~M., \& Dullemond, C.~P.\ 2008, \aap, 486, 245 
\bibitem[Cutri et al.(2003)]{2003tmc..book.....C} Cutri, R.~M., Skrutskie, M.~F., van Dyk, S., et al.\ 2003, The IRSA 2MASS All-Sky Point Source Catalog, NASA/IPAC Infrared Science Archive  
\bibitem[Davis et al.(2007)]{2007MNRAS.374...29D} Davis, C.~J., Kumar, M.~S.~N., Sandell, G., et al.\ 2007, \mnras, 374, 29 
\bibitem[Deharveng et al.(2010)]{2010A&A...523A...6D}Deharveng, L., Schuller, F., Anderson, L.~D., Zavagno, A., Wyrowski, F., Menten, K.~M., Bronfman, L., Testi, L., Walmsley, C.~M. \& Wienen, M.\ 2010 \aap, 523, 6
\bibitem[Deharveng et al.(2012)]{2012A&A...546A..74D} Deharveng, L., Zavagno, A., Anderson, L.~D., Motte, F., Abergel, A., Andr\'e, P., Bontemps, S., Leleu, G., Roussel, H. and Russeil, D. 2012 \aap, 546, 74.
\bibitem[Dickel et al.(1969)]{1969A&A.....1..270D} Dickel, H.~R., Wendker, H., \& Bieritz, J.~H.\ 1969, \aap, 1, 270 
\bibitem[Enoch et al.(2009)]{2009ApJ...692..973E} Enoch, M.~L., Evans, N.~J., II, Sargent, A.~I., \& Glenn, J.\ 2009, \apj, 692, 973 
\bibitem[Evans et al.(2009)]{2009ApJS..181..321E} Evans, N.~J., II, Dunham, M.~M., J{\o}rgensen, J.~K., et al.\ 2009, \apjs, 181, 321 
\bibitem[Fazio et al.(2004)]{2004ApJS..154...10F} Fazio, G.~G., Hora, J.~L., Allen, L.~E., et al.\ 2004, \apjs, 154, 10 
\bibitem[Fischer et al.(2010)]{2010A&A...518L.122F} Fischer, W.~J., Megeath, S.~T., Ali, B., Tobin, J.~J., Osorio, M., Allen, L.~E., Kryukova, E., Stanke, T., Stutz, A.~M., Bergin, E., Calvet, N., di Francesco, J., Furlan, E. Hartmann, L., Henning, T., Krause, O., Manoj, P., Maret, S., Muzerolle, J., Myers, P., Neufeld, D., Pontoppidan, K., Poteet, C.~A., Watson, D.~M. Wilson, T. 2010 \aap 518, 122
\bibitem[Greene et al.(1994)]{1994ApJ...434..614G} Greene, T.~P., Wilking, B.~A., Andre, P., Young, E.~T., \& Lada, C.~J.\ 1994, \apj, 434, 614 
\bibitem[Gutermuth et al.(2011)]{2011ApJ...739...84G} Gutermuth, R.~A., Pipher, J.~L., Megeath, S.~T., et al.\ 2011, \apj, 739, 84 
\bibitem[Gutermuth et al.(2009)]{2009ApJS..184...18G} Gutermuth, R.~A., Megeath, S.~T., Myers, P.~C., et al.\ 2009, \apjs, 184, 18
\bibitem[Gutermuth et al.(2008)]{2008ApJ...674..336G} Gutermuth, R.~A., et al.\ 2008, \apj, 674, 336 
\bibitem[Hanson(2003)]{2003ApJ...597..957H} Hanson, M.~M.\ 2003, \apj, 597, 957 
\bibitem[Hartmann et al.(1997)]{1997ApJ...475..770H} Hartmann, L., Cassen, P., \& Kenyon, S.~J.\ 1997, \apj, 475, 770 
\bibitem[Harvey et al.(2007)]{2007ApJ...663.1149H} Harvey, P., Mer{\'{\i}}n, B., Huard, T.~L., et al.\ 2007, \apj, 663, 1149  
\bibitem[Hennemann et al.(2012)]{2012A&A...543L...3H} Hennemann, M., Motte, F., Schneider, N., et al.\ 2012, \aap, 543, L3 
\bibitem[Hsu et al.(2012)]{2012ApJ...752...59H} Hsu, W.-H., Hartmann, L., Allen, L., et al.\ 2012, \apj, 752, 59 
\bibitem[Hsu et al.(2013)]{2013ApJ...764..114H} Hsu, W.-H., Hartmann, L., Allen, L., et al.\ 2012, \apj, 764, 114 
\bibitem[Kn{\"o}dlseder(2000)]{2000A&A...360..539K} Kn{\"o}dlseder, J.\ 2000, \aap, 360, 539
\bibitem[Krumholz et al.(2010)]{2010ApJ...713.1120K} Krumholz, M.~R., Cunningham, A.~J., Klein, R.~I., \& McKee, C.~F.\ 2010, \apj, 713, 1120 
\bibitem[Kryukova et al.(2012)]{2012AJ....144...31K} Kryukova, E., Megeath, S.~T., Gutermuth, R.~A., et al.\ 2012, \aj, 144, 31
\bibitem[Lada \& Wilking(1984)]{1984ApJ...287..610L} Lada, C.~J., \& Wilking, B.~A.\ 1984, \apj, 287, 610 
\bibitem[Landecker(1984)]{1984AJ.....89...95L} Landecker, T.~L.\ 1984, \aj, 89, 95 
\bibitem[Lonsdale et al.(2003)]{2003PASP..115..897L} Lonsdale, C.~J., Smith, H.~E., Rowan-Robinson, M., et al.\ 2003, \pasp, 115, 897 
\bibitem[Lucas et al.(2008)]{2008MNRAS.391..136L} Lucas, P.~W., Hoare, M.~G., Longmore, A., et al.\ 2008, \mnras, 391, 136 
\bibitem[McClure et al.(2010)]{2010ApJS..188...75M} McClure, M.~K., Furlan, E., Manoj, P., et al.\ 2010, \apjs, 188, 75 
\bibitem[McKee \& Tan(2003)]{2003ApJ...585..850M} McKee, C.~F., \& Tan, J.~C.\ 2003, \apj, 585, 850
\bibitem[Massey \& Thompson(1991)]{1991AJ....101.1408M} Massey, P., \& Thompson, A.~B.\ 1991, \aj, 101, 1408 
\bibitem[Megeath \& Wilson(1997)]{1997AJ....114.1106M} Megeath, S. T. \& Wilson, T.~L.\ 1997, \aj 114, 1106
\bibitem[Megeath et al.(2004)]{2004ApJS..154..367M} Megeath, S.~T., Allen, L.~E., Gutermuth, R.~A., et al.\ 2004, \apjs, 154, 367 
\bibitem[Megeath et al.(2009)]{2009AJ....137.4072M} Megeath, S.~T., Allgaier, E., Young, E., et al.\ 2009, \aj, 137, 4072 
\bibitem[Megeath et al.~(2012)]{2012AJ....144..192M} Megeath, S.~T., Gutermuth, R., Muzerolle, J., Kryukova, E., Flaherty, K., Hora, J., Allen, L.~E., Hartmann, L., Myers, P.~C., Pipher, J.~L., Stauffer, J., Young, E.~T., Fazio, G.~G., et al. \ 2012,  \aj, 144, 192 
\bibitem[Motte et al.(2007)]{2007A&A...476.1243M} Motte, F., Bontemps, S., Schilke, P., et al.\ 2007, \aap, 476, 1243 
\bibitem[Muzerolle et al.(2004)]{2004ApJS..154..379M} Muzerolle, J., Megeath, S.~T., Gutermuth, R.~A., et al.\ 2004, \apjs, 154, 379
\bibitem[Offner \& McKee(2011)]{2011ApJ...736...53O} Offner, S.~S.~R., \& McKee, C.~F.\ 2011, \apj, 736, 53 
\bibitem[Palla \& Stahler(1993)]{1993ApJ...418..414P} Palla, F., \& Stahler, S.~W.\ 1993, \apj, 418, 414 
\bibitem[Polychroni(2013)]{2013arXiv1309.2332P} Polychroni, D., Schisano, E., Elia, D., Roy, A., Molinari, S., 
Martin, P., Andre, P.,  Turrini, D., Rygl, K.~L.~J., Benedettini, M., Busquet, G., di Giorgio, A.~M., Pestalozzi, M., Pezzuto, S., Arzoumanian, D., Bontemps, S., Di Francesco, J., Hennemann, M., Hill, T., Konyves, V., Menshchikov, A., Motte, F., Nguyen-Luong, Q., Peretto, N., Schneider, N. and 
	White, G. 2013arXiv1309.2332P
\bibitem[Rieke et al.(2004)]{2004ApJS..154...25R} Rieke, G.~H., Young, E.~T., Engelbracht, C.~W., et al.\ 2004, \apjs, 154, 25 
\bibitem[Reipurth \& Schneider(2008)]{2008hsf1.book...36R} Reipurth, B., \& Schneider, N.\ 2008, Handbook of Star Forming Regions, Volume I, 36 
\bibitem[Robitaille et al.(2006)]{2006ApJS..167..256R} Robitaille, T.~P., Whitney, B.~A., Indebetouw, R., Wood, K., \& Denzmore, P.\ 2006, \apjs, 167, 256 
\bibitem[Rygl et al.(2012)]{2012A&A...539A..79R} Rygl, K.~L.~J., Brunthaler, A., Sanna, A., et al.\ 2012, \aap, 539, A79 
\bibitem[Schneider et al.(2012)]{2012A&A...542L..18S} Schneider, N., G{\"u}sten, R., Tremblin, P., et al.\ 2012, \aap, 542, L18  
\bibitem[Schneider et al.(2011)]{2011A&A...529A...1S} Schneider, N., Bontemps, S., Simon, R., et al.\ 2011, \aap, 529, A1
\bibitem[Schneider et al.(2010)]{2010A&A...520A..49S} Schneider, N.,  Csengeri, T., Bontemps, S., Motte, F. Simon, R., Hennebelle P., Federrath, C. \& Klessen, R., \aap, 520, A49
\bibitem[Schneider et al.(2007)]{2007A&A...474..873S} Schneider, N., Simon, R., Bontemps, S., Comer{\'o}n, F., \& Motte, F.\ 2007, \aap, 474, 873 
\bibitem[Schneider et al.(2006)]{2006A&A...458..855S} Schneider, N., Bontemps, S., Simon, R., et al.\ 2006, \aap, 458, 855 
\bibitem[Skrutskie et al.(2006)]{2006AJ....131.1163S} Skrutskie, M.~F., Cutri, R.~M., Stiening, R., et al.\ 2006, \aj, 131, 1163 
\bibitem[Stutz et al.(2013)]{2013ApJ...767...36S}, Stutz, A.~M., Tobin, J.~J., Stanke, T., Megeath, S.~T., Fischer, W.~J., Robitaille, T., Henning, T. Ali, B., di Francesco, J., Furlan, E., Hartmann, L., Osorio, M., Wilson, T.~L., Allen, L., Krause, O. and Manoj, P. 2013 \apj, 767, 36.
\bibitem[Whitney et al.(2004)]{2004ApJ...617.1177W} Whitney, B.~A., Indebetouw, R., Bjorkman, J.~E., \& Wood, K.\ 2004, \apj, 617, 1177 
\bibitem[Wilson et al.(2005)]{2005A&A...430..523W} Wilson, B.~A., Dame, T.~M., Masheder, M.~R.~W., \& Thaddeus, P.\ 2005, \aap, 430, 523 
\bibitem[Winston et al.(2007)]{2007ApJ...669..493W} Winston, E., Megeath, S.~T., Wolk, S.~J., et al.\ 2007, \apj, 669, 493
\bibitem[Wright et al.(2012b)]{2012arXiv1208.0211W} Wright, N.~J., Bouy, H., Drake, J.~J., et al.\ 2012, arXiv:1208.0211
\bibitem[Wright et al.(2012a)]{2012ApJ...746L..21W} Wright, N.~J., Drake, J.~J., Drew, J.~E., et al.\ 2012, \apjl, 746, L21 
\bibitem[Wright et al.(2010)]{2010ApJ...713..871W} Wright, N.~J., Drake, J.~J., Drew, J.~E., \& Vink, J.~S.\ 2010, \apj, 713, 871 
\bibitem[Zapata et al.(2012)]{2012ApJ...744...86Z} Zapata, L.~A., Loinard, L., Su, Y.-N., et al.\ 2012, \apj, 744, 86
\end{thebibliography}
\end{document}